\documentclass[showpacs,preprintnumbers,amsmath,amssymb,aps, prd]{revtex4}
\usepackage{graphicx}
\usepackage{grffile}
\usepackage{mathrsfs, mathtools}
\usepackage{caption}
\usepackage{float}
\usepackage{xcolor}
\usepackage{changepage}
\usepackage{dcolumn}
\usepackage{bm}
\usepackage{graphicx,subfigure,epsfig}
\usepackage{multirow}
\newcommand{\lagr}{\mathscr{L}}

\def\a{\alpha}
\def\b{\beta}
\def\f{\frac}
\def\e{\mathcal{E}}
\def\l{\mathcal{L}}
\def\c{\cite}
\def\r{\ref}
\def\fr{f(r)}
\def\gr{g(r)}
\def\hr{h(r)}
\def\s{Schwarzschild }
\newcommand\be{\begin{equation}}
\newcommand\ee{\end{equation}}
\newcommand\ba{\begin{eqnarray}}
\newcommand\ea{\end{eqnarray}}
\newcommand\nn{\nonumber}
\newcommand\lt{\left}
\newcommand\rt{\right}
\newcommand\pt{\partial}
\newcommand\tx{\text}

\begin{document}
\title{Observational signature of Lorentz violation in Kalb-Ramond field model and Bumblebee model: A comprehensive comparative study}
\author{Sohan Kumar Jha}
\email{sohan00slg@gmail.com}
\affiliation{Chandernagore College, Chandernagore, Hooghly, West
Bengal, India}

\date{\today}
\begin{abstract}
\begin{center}
Abstract
\end{center}
This article is devoted to the comparative study of the effects of the Lorentz symmetry violation (LV) arising in Kalb-Ramond (KR) and Bumblebee (BM) field models. We study optical appearance with accretion, Quasinormal modes, ringdown waveforms, Hawking radiation, and weak gravitational lensing for both black holes (BHs) and compare their observational impacts. The horizon radius, photon radius, and the critical impact parameter for KR BHs decrease with the LV parameter $\a$. In contrast, they remain independent of the BM parameter $\b$ and have values the same as those for \s BH. Their qualitative and quantitative variation with $\a$ are illustrated for KR BH. Our study reveals that the observed intensity for static and infalling accretions peaks at the critical impact parameter. We find that a KR BH is brighter than a \s or BM BH for static as well as infalling accretion. The BM BH, on the other hand, is brighter than a \s BH when the accretion is static but becomes darker for an infalling accretion. The shadow's size, independent of the accretion type, is smaller for a KR BH. Our investigation into quasinormal modes (QNMs) with the help of the $6$th order Pad\'{e} averaged WKB method and ringdown waveforms provide deeper insight into the difference in observational imprints of LV parameters. It reveals that GWs emitted by KR BHs have larger frequencies and decay faster than those emitted by \s or BM BHs for scalar and electromagnetic perturbations. GWs emitted by BM BHs have lower frequency and decay slower than those emitted by \s BHs for scalar perturbation, whereas the frequency is higher for electromagnetic perturbation. We then study the greybody factor (GF) and power emitted for both BHs. The Hawking temperature is higher for a KR BM and lower for a BM BH than a \s BH. It also reveals that the transmission probability decreases with $\a$ and $\b$. A comparison of GFs for KR and BM BHs reveals that the transmission probability is higher for BM BH. We also study the effect of LV on the power emitted in the form of Hawking radiation. Power received by an asymptotic observer is larger for a KR BH. We finally try to differentiate between KR and BM BHs based on their observational signatures in the weak gravitational lensing. We obtain higher-order corrections in the deflection angle and graphically illustrate the impact of $\a$ and $\b$. We observe that a light ray gets deflected most from its path when passing by a \s BH, and the deflection is least when it passes by a KR BH. Our study conclusively shows that we can differentiate between KR and BM BHs based on astrophysical observations.
\end{abstract}
\maketitle
\section{Introduction}
The Lorentz symmetry is the central pillar of the two most successful theories: general relativity and the standard model of particle physics. Despite their astounding success in predicting various experimental observations, it is widely believed that the Lorentz symmetry may be violated at higher energies in various theories, such as string theory, non-commutative field theory, and so on [\citenum{Kostelecky1989a}-\citenum{Cohen2006}]. The Lorentz symmetry can either be broken spontaneously or explicitly. When broken explicitly, the Lagrange density no longer remains invariant under the Lorentz transformation, and we have physical laws taking different forms in certain reference frames. When the Lorentz symmetry is spontaneously broken, although the Lagrange density shows Lorentz invariance, the Lorentz symmetry is broken in the ground state. The standard model extension provides a general framework for incorporating LV \cite{Kostelecky2004a}. Bumblebee models within this extension are one of the simplest theories where LV occurs spontaneously due to the non-zero vacuum expectation value of a vector field called bumblebee field \cite{Kostelecky1989a, Kostelecky1989, Kostelecky1989b, Bailey2006, Bluhm2008a}. Casana et al., in their article \c{bm}, reported a \s-like solution. Various aspects of the BH, such as lensing, accretion, etc., have extensively been studied in [\citenum{Ovgun2018}-\citenum{Liang2023}]. Some other notable articles related to BHs in the bumblebee model are [\citenum{Ding2020a}-\citenum{Amarilo2023}]. In the KR model, LV is spontaneously broken due to  a rank-two antisymmetric tensor field known as the Kalb-Ramond field assumes non-zero vacuum expectation value \c{Altschul2010} . Various properties of the KR field have been studied in [\citenum{Kao1996}-\citenum{Chakraborty2016}]. Lessa et al. obtained an exact non-rotating solution in the KR model \cite{Lessa2020}. For more details regarding different aspects of Lessa solution and rotating KR BHs, please see [\citenum{Atamurotov2022}-\citenum{Maluf2022a}]. A different exact static and spherically symmetric solution in the KR field model is obtained by Ke Yang et al. in \c{kr}. This solution will be the subject of our study along with the solution given in \c{bm}.\\
BHs acquire mass (and possibly angular momentum) through the accretion process by capturing the ambient matter surrounding BHs \c{IG}. The kinetic energy of accreting matter increases at the cost of its gravitational energy, which enhances the energy radiated by the central object. The accretion process is believed to be the primary energy source of astronomical massive objects such as quasars and active galactic nuclei. It helps us explain astrophysical phenomena such as gamma-ray bursts and tidal disruption events. Exploring differences in the impact LV in the BM model and LV in the KR model have on the shadow with static and infalling accretion is our primary goal here. Shadow of a BH encodes important information regarding the central object. It directly manifests the strong gravitational field of BHs. Since one can find the observational signature of LV in the shadow and accretion, it would be instructive to study the shadow in the presence of accretion. Please see articles [\citenum{RN}-\citenum{XX}] for more details.\\
The emergence of QNMs is a result of perturbation of BHs. These modes are transient in nature. BH, on getting perturbed, emits gravitational waves (GWs), and eventually, the perturbed state of the BH dies down. There are three phases that a BH, on perturbation, experiences \c{KONOR}: initially, there is an outburst of radiation depending on the initial condition, then we have the ringdown phase, and the final one is the tail where the perturbation dies down by a power or exponential law. QNMs are associated with the ringdown phase. They are functions of the model parameters characterizing the central object and the type of perturbation BH encounters. They do not depend on the initial conditions, signifying that the BH continually interacts with the field responsible for the perturbation. Thus, probing BH perturbation and QNMs would be imperative to gauge the impact of LV and to differentiate between two BHs that incorporate the LV effect on the basis of an observational signature. The prospect of QNMs getting detected by LIGO, VIRGO, and LISA in the future makes it more pertinent to study QNMs \c{LIGO1, LIGO2, LIGO3, LIGO4, VIRGO}. A great deal of research has been carried out to understand QNMs for different BH metrics [\citenum{13}-\citenum{50}].\\
Hawking radiation and the association of temperature with BH horizon are fascinating subjects that have intrigued researchers for quite some time. Hawking showed that a BH emits radiation by considering quantum consequences \c{HAWKING}. It can be explained with the help of a pair production near the horizon where one of the particles gets consumed by the BH, whereas the second particle moves towards an asymptotic observer. It is the second particle that constitutes the Hawking radiation [\citenum{HH}-\citenum{HC}]. It was later shown by Bekenstein and Keif that to be consistent with thermodynamics, a temperature needs to be associated with the BH horizon \c{BEK, KEIF}. We can obtain Hawking temperature using different techniques [\citenum{SW}-\citenum{SI}]. The greybody factor (GF) is a significant quantity that provides the probability of Hawking radiation getting detected by a distant observer. Out of various methods at our disposal to bound GF[\citenum{qn32}-\citenum{WJ}], we employed the elegant technique illustrated in \c{GB, GB1, GB2}. Here, Hawking radiation is also investigated to differentiate between KE and BM BHs on the basis of LV imprints in observation. We also explore the weak gravitational lensing (GL) due to its dependence on the nature of the underlying spacetime. In gravitational lensing, a massive central object acting as a lens deflects light rays from its path, where the deflection angle is a function of model parameters. It has been extensively studied to probe dark matter and energy and also for the determination of the mass of galaxies [\citenum{251}-\citenum{271}]. GL can be studied in the strong field limit and weak field limit. It is the latter we are interested in. The central pillar of studies made in the weak GL is the seminal work by Gibbons and Werner \c{GW}. This technique was subsequently extended to stationary spacetimes in \c{WERNER} and to finite distances in \c{ISHIHARA1, ISHIHARA2}. These studies form the basis of further research in \c{ONO1, ONO2, ONO3}. Higher order corrections in the deflection angle were later obtained in \cite{CRISNEJO}. Please refer to the articles for studies in different spacetimes [\citenum{411}-\citenum{511}]. We will explore weak GL to find differential observational signatures of LV arising in KR and BM models.\\
We compose the article in the following manner. Sec. II briefly introduces KR and BM metrics, and unstable circular orbits are studied. Sec. III is where the optical appearance of BH with accretion is examined, and a qualitative comparison is carried out to differentiate between KR and BM BHs. Sec. IV and V are devoted to studying differences in the implication of LV on QNMs in KR and BM models. Differential observational signatures of LV in KR and BM models with respect to Hawking radiation are explored in Sec. VI. We try to differentiate between KR and BM BHs based on the deflection angle in the weak field limit in Sec. VII. We finally elaborate on our findings in Sec. VIII.
\section{Static and spherically symmetric BH solutions with Lorentz symmetry violation}
We introduce static and spherically symmetric solutions resulting from KR and BM. In both models, the the non-zero vacuum expectation value of a field breaks the Lorentz symmetry spontaneously : it is a vector field in BM and a tensor field in KR. The static and spherically symmetric solution in the BM model is \c{bm}
\ba
ds^2&=&-\fr dt^2+\f{dr^2}{\gr}+\hr\lt(d\theta^2+sin^2\theta d\phi^2\rt) \\
&=& -\lt(1-\f{2M}{r}\rt) dt^2+\f{1+\b}{\f{1}{1-\a}-\f{2M}{r}}dr^2+ r^2\lt(d\theta^2+sin^2\theta d\phi^2\rt),
\label{bmmetric}
\ea
where $\fr=1-\f{2M}{r}$, $\gr=\f{1-\f{2M}{r}}{1+\b}$, and $\hr=r^2$. The BH solution obtained from the KR model is \c{kr}
\ba
ds^2&=&-\fr dt^2+\f{dr^2}{\gr}+\hr\lt(d\theta^2+sin^2\theta d\phi^2\rt)\\
&=& -\lt(\f{1}{1-\a}-\f{2M}{r}\rt) dt^2+\f{dr^2}{\f{1}{1-\a}-\f{2M}{r}}+ r^2\lt(d\theta^2+sin^2\theta d\phi^2\rt),
\label{krmetric}
\ea
where $\fr=\gr=\f{1}{1-\a}-\f{2M}{r}$ and $\hr=r^2$. The KR(BM) model reduces to \s solution in the limit $\a(\b)\rightarrow 0$. The Kretschmann scalars for the KR BH ($K_\a$) and BM BH ($K_\b$) are given by
\ba\nn
K_\a=\f{48 M^2}{r^6}-\f{16 \a M}{(1-\a) r^5}+\f{4 \a^2}{(\a-1)^2 r^4},\\
K_\b=\f{4}{\lt(1+\b\rt)^2}\lt(\f{12M^2}{r^6}+\f{4\b M}{r^5}+\f{\b^2}{r^4}\rt).
\ea
Since the Kretschmann scalars for KR and BM BHs differ from that for a \s BH, there exists no coordinate transformation that can eliminate LV from metrics [$2$] and [$4$]. A KR BH has an event horizon at $r_h=2M(1-\a)$, which depends on the parameter $\a$, whereas the horizon radius for a BM BH is $r_h=2M$, irrespective of the value of $\b$. To observe shadow, we must consider the motion of massless particles in the background of the BH. Since the BH solutions under consideration are spherically symmetric, we may consider null geodesics confined to the equatorial plane. As such, the Lagrangian corresponding to the metric [$1$] is
\be
\lagr=\f{1}{2}\lt(-\fr \dot{t}^2+\f{\dot{r}^2}{\gr}+\hr \dot{\phi}^2\rt).
\ee
Here, $\dot{}$ is the differentiation with respect to the affine parameter $\lambda$. Owing to the static and spherically symmetric nature of the spacetime, we have two conserved quantities: energy $\e=-p_{t}=-\f{\pt \lagr}{\pt \dot{t}}=\fr \dot{t}$ and angular momentum $\l=p_{\phi}=\f{\pt \lagr}{\pt \dot{\phi}}=\hr \dot{\phi}$. With these conjoined with the fact that $\lagr=0$ for null geodesics, we obtain
\ba\nn
&&\dot{r}^2+\lt(-\e^2 \f{\gr}{\fr}+\l^2 {\gr}{\hr}\rt)=0\\
&\Rightarrow& \dot{r}^2+V_{eff}(r)=0,
\label{rdot}
\ea
where $V_{eff}=-\e^2 \f{\gr}{\fr}+\l^2 {\gr}{\hr}$ is the effective potential. Following are the conditions for an unstable spherical orbit at radius $r_p$:
\be
V_{eff}(r_p)=0, \quad \f{\pt V_{eff}}{\pt r}|_{r=r_p}=0, \quad \tx{and} \quad \f{\pt^2 V_{eff}}{\pt r^2}|_{r=r_p}<0.
\ee
The above conditions yield $\fr' \hr=\hr' \fr$ evaluted at $r=r_p$, where $'$ implies differentiation with respect to $r$. For a KR BH, the position of the unstable spherical orbit is $r_p=3M(1-\a)$, and the spherical radius for a BM BH is $r_p=3M$. Thus, LV does not affect the photon orbit in the case of a BM BH. However, due to the presence of LV, the spherical radius decreases with $\alpha$. The critical impact parameter associated with the spherical orbit is
\be
b_p=\sqrt{\f{\hr}{\fr}}.
\ee
For a KR BH, $b_p=3\sqrt{3}\lt(1-\a\rt)^{3/2}M$ and for a BM BH, $b_p=3\sqrt{3}M$. Similar to the photon radius, increasing the KR parameter $\alpha$ would decrease the impact parameter. Although LV leaves an imprint on the impact parameter in KR, it is absent in BM. This is because the LV parameter in BM appears in the radial component of the metric. In contrast, the spherical radius and the impact parameter depend on the metric's time and the azimuthal components.
\begin{figure}[H]
\centering
\subfigure[]{
\label{vfig1}
\includegraphics[width=0.4\columnwidth]{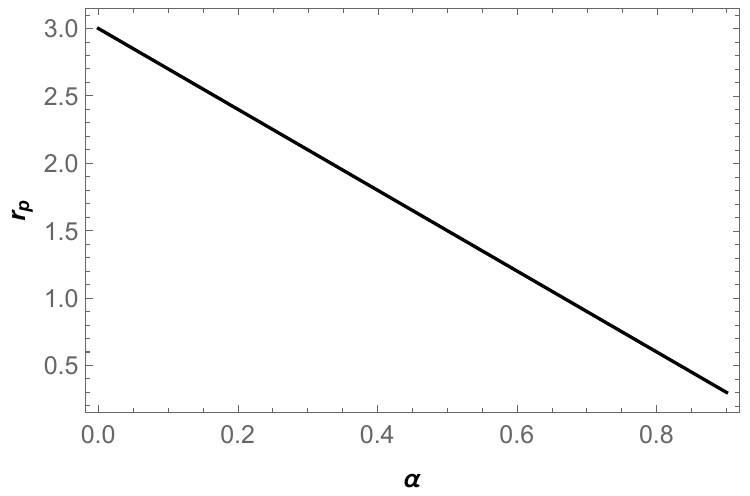}
}
\subfigure[]{
\label{vfig2}
\includegraphics[width=0.4\columnwidth]{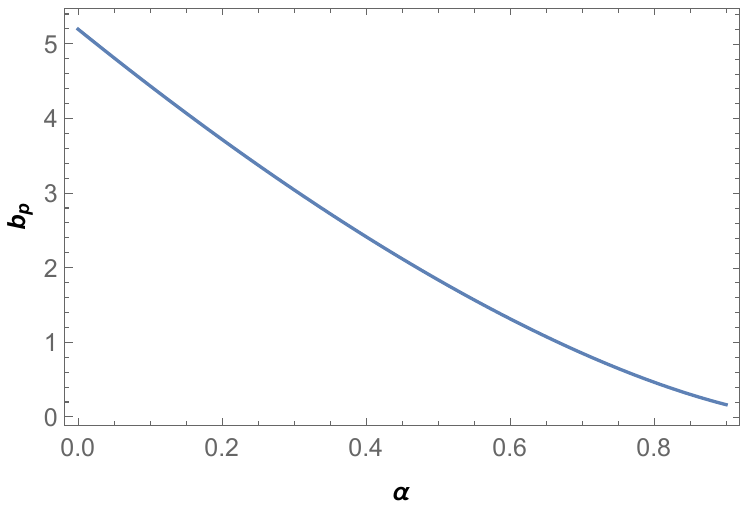}
}
\caption{Variation of $r_p$ and $b_p$ against $\alpha$.}
\label{radius}
\end{figure}
\begin{table}[!htp]
\centering
\caption{Various values of horizon radius, photon radius, and impact parameter for different values of $\alpha$.}
\begin{tabular}{|c|c|c|c|c|c|c|c|c|c|c|c|}
\hline
$\alpha$ & 0. & 0.1 & 0.2 & 0.3 & 0.4 & 0.5 & 0.6 & 0.7 & 0.8 & 0.9 \\
\hline
$r_{\text{eh}}$ & 2. & 1.8 & 1.6 & 1.4 & 1.2 & 1. & 0.8 & 0.6 & 0.4 & 0.2 \\
\hline
$r_p$ & 3.0 & 2.7 & 2.4 & 2.1 & 1.8 & 1.5 & 1.2 & 0.9 & 0.6 & 0.3 \\
\hline
$b_p$ & 5.19615 & 4.43655 & 3.71806 & 3.04319 & 2.41495 & 1.83712 & 1.31453 & 0.853815 & 0.464758 & 0.164317 \\
\hline
\end{tabular}
\label{shadowradius}
\end{table}
The qualitative variation of the photon radius and the critical impact parameter are shown in Fig. [\r{radius}], and their quantitative nature of variation is shown in Table [\r{shadowradius}]. The significant impact of KR parameter $\a$ on these observables is evident from above. Next, we will explore LV's impact on the optical appearance of BH with different accretion models.
\section{Black hole shadow with accretion}
Accretion is the process whereby massive astronomical objects gain mass by capturing surrounding matter. It is believed to be the primary energy source of many massive objects, such as quasars and active galactic nuclei. The existence of an accretion disk around a BH significantly impacts its optical appearance. Since the accretion and the BH shadow largely depend on the nature of the underlying spacetime, it would be instructive to study the BH shadow with accretion. Static and infalling accretion will be considered in our study.
\subsection{Shadow with static accretion}
The static accretion disk is considered to be optically and geometrically thin. The measured intensity by an asymptotic observer is \cite{mj97, bambi13}
\begin{equation}
\label{intensity}
I(\nu_{obs})=\int_{ray}g_{f}^{3}j(\nu_{em})dl_{prop},
\end{equation}
where $\nu_{obs}$ and $\nu_{em}$ are, respectively, the observed and emitted frequencies, the redshift factor $g_{f}=\sqrt{\fr}$, $j(\nu_{em})$ is the emissivity per unit volume, and the proper differential length $dl_{prop}=\sqrt{\f{1}{\gr}+\hr\Big(\f{d\phi}{dr}\Big)^{2}}dr$. We need to obtain the expression for $\f{d\phi}{dr}$.
\begin{center}
\includegraphics[width=6cm,height=6cm]{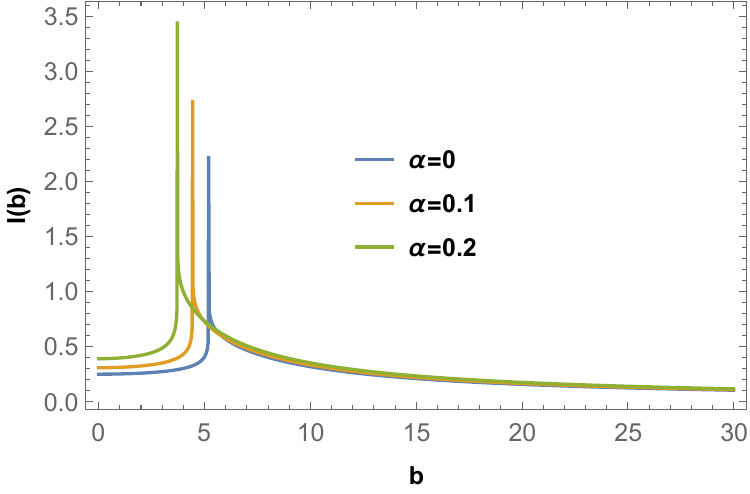}
\includegraphics[width=6cm,height=6cm]{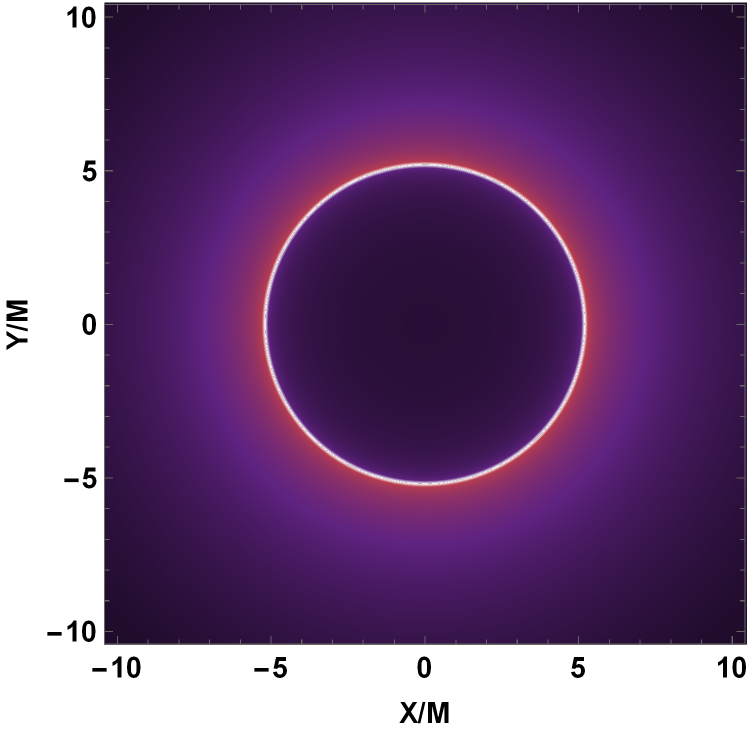}
\includegraphics[width=0.5cm,height=6cm]{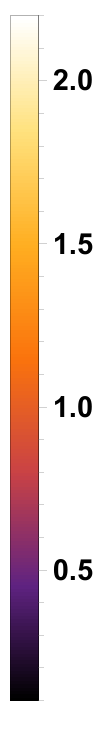}
\includegraphics[width=6cm,height=6cm]{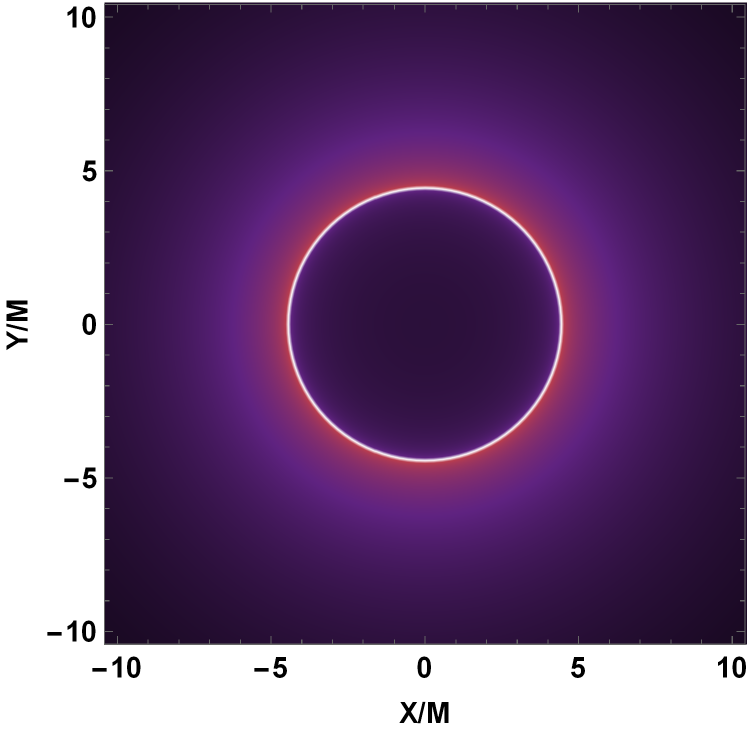}
\includegraphics[width=0.5cm,height=6cm]{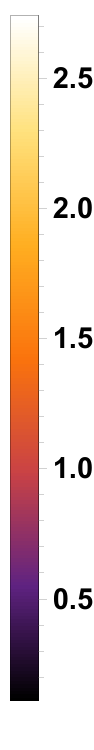}
\includegraphics[width=6cm,height=6cm]{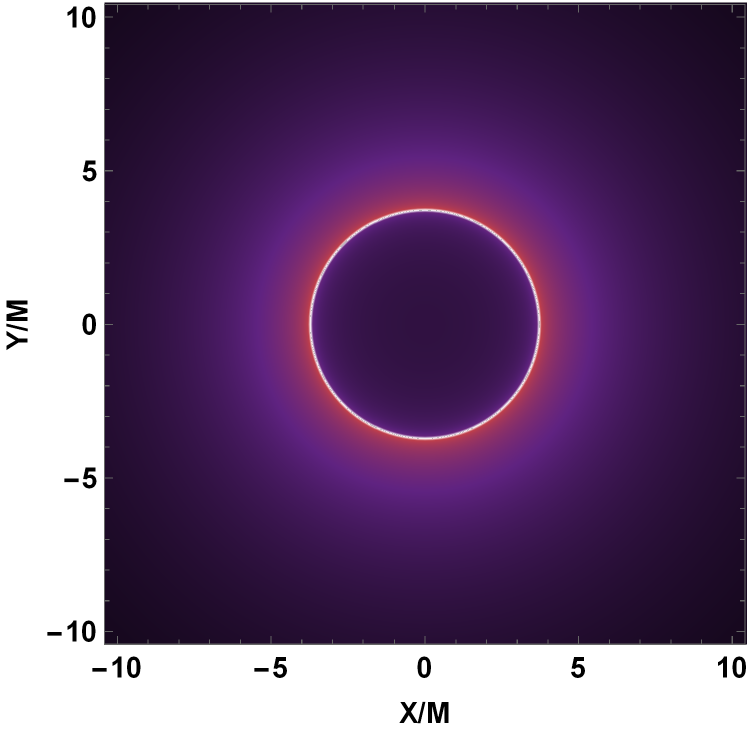}
\includegraphics[width=0.5cm,height=6cm]{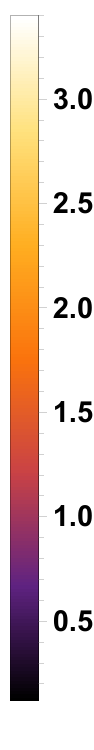}
\parbox[c]{15.0cm}{\footnotesize{\bf Fig~2.} 
Observed intensity and black hole shadows with static spherical accretion for KR BH. The upper right is for $\a=0$, the lower left one is for $\a=0.1$, and the lower right one is for $\a=0.2$.}
\label{statickr}
\end{center}
To this end, we use the Eq. [\r{rdot}] conjoined with $\dot{\phi}^2=\f{\l^2}{\hr^2}$. This yields
\be
dl_{prop}=\sqrt{\f{\hr}{\gr(\hr-b^2 \fr)}}
\ee
With the assumption that the emission is monochromatic, we take $j(\nu_{em})\propto \f{\delta(\nu_{em}-\nu_{t})}{r^2}$, $\nu_{t}$ being the rest-frame frequency. Thus, the expression for the measured intensity becomes
\begin{equation}
\label{intensity}
I(\nu_{obs})=\int_{ray}\f{\fr^{3/2}}{r^2}\sqrt{\f{\hr}{\gr(\hr-b^2 \fr)}}.
\end{equation}
Photons with an impact parameter equal to their critical value get trapped in the unstable circular orbit. If they get perturbed from their circular path, they may either move towards the BH and get swallowed or move away from it and reach the distant observer. The latter photons are responsible for the bright circular ring seen in Figs. [$2$, $3$]. Photons with impact parameter $b<b_{p}$ come too close to the BH to escape its strong gravitational pull and get consumed by it. Those photons with $b>b_{p}$ get deflected towards the asymptotic observer.
\begin{center}
\includegraphics[width=6cm,height=6cm]{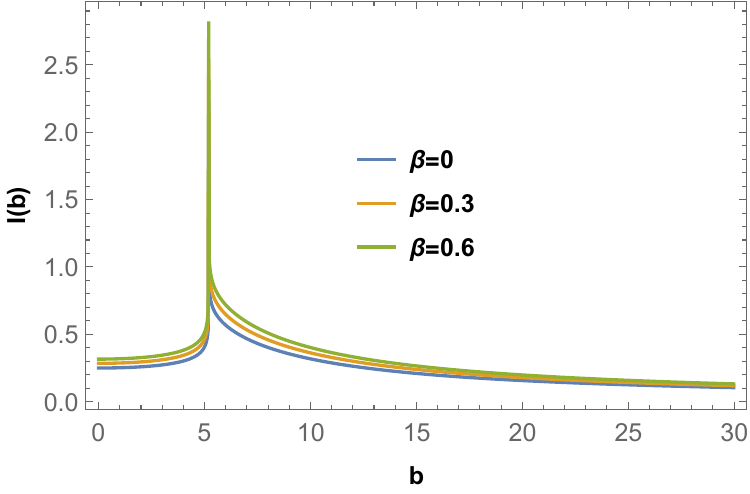}
\includegraphics[width=6cm,height=6cm]{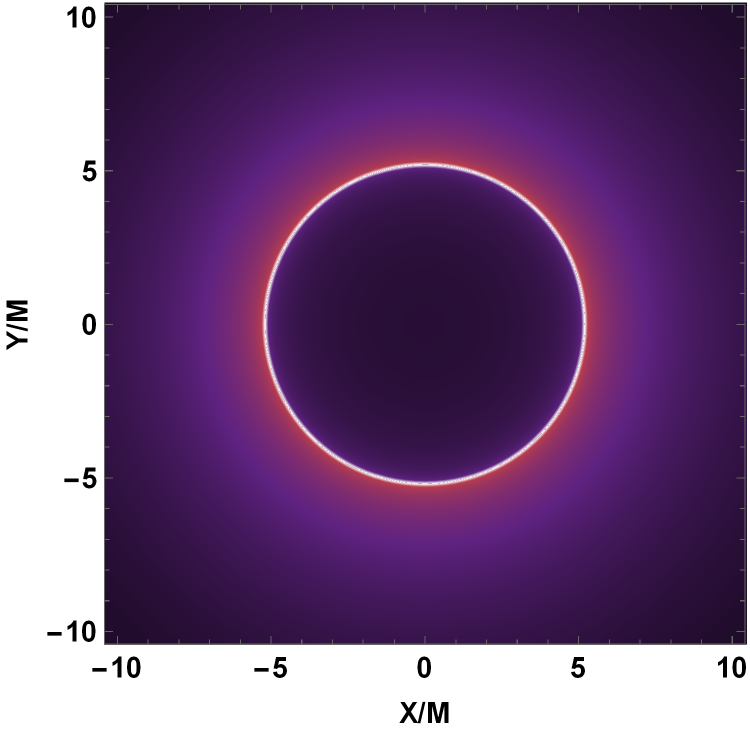}
\includegraphics[width=0.5cm,height=6cm]{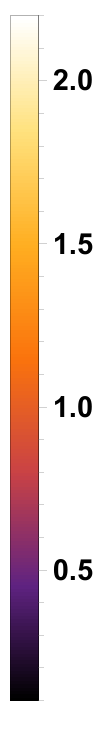}
\includegraphics[width=6cm,height=6cm]{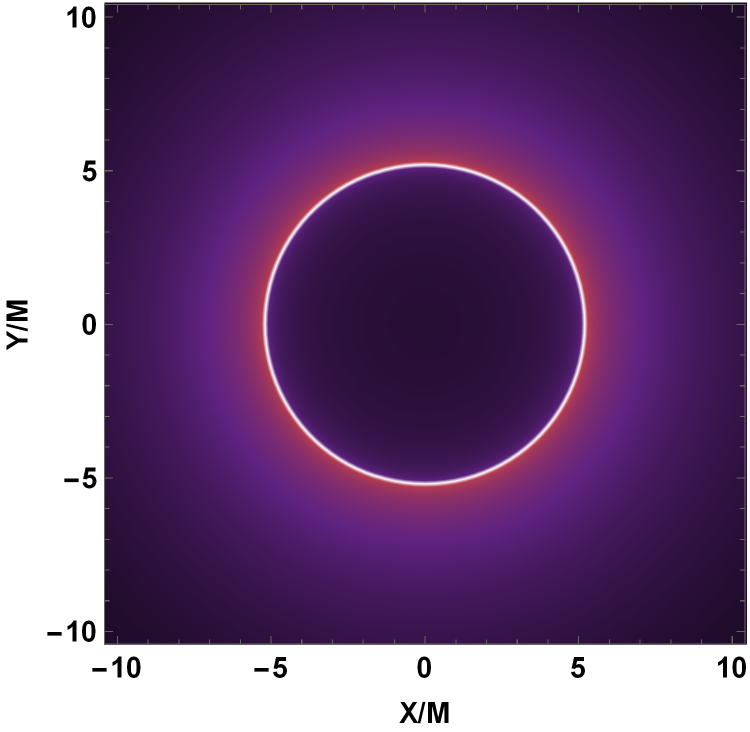}
\includegraphics[width=0.5cm,height=6cm]{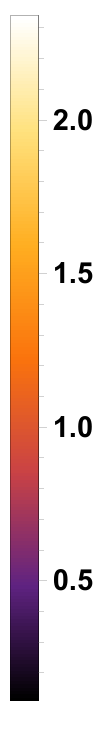}
\includegraphics[width=6cm,height=6cm]{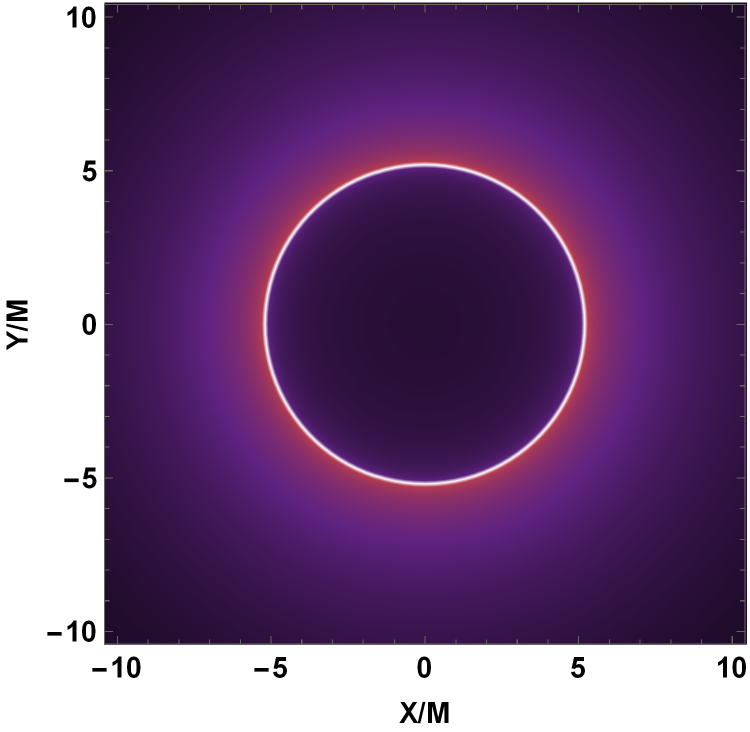}
\includegraphics[width=0.5cm,height=6cm]{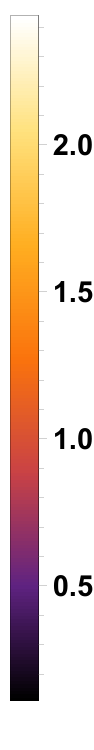}
\parbox[c]{15.0cm}{\footnotesize{\bf Fig~3.} 
Observed intensity and black hole shadows with static spherical accretion BM BH. The upper right is for $\b=0$, the lower left one is for $\b=0.1$, and the lower right one is for $\b=0.2$.}
\label{staticbm}
\end{center}
It is evident from Figs. [$2$, $3$] that the intensity initially increases with the impact parameter before reaching its peak value at $b_{p}$ and then sharply decreases with $b$. The peak position remains unaltered, and the peak value changes marginally with the BM parameter $\b$. However, the position of the peak shifts towards the left, and the maximum intensity increases significantly with increasing KR parameter $\a$. It is because the intensity reaches its maximum value at $b=b_p$, which does not change with $\b$ but decreases with $\a$. The region within the shadow is not entirely black since a small fraction of photons can escape a BH. It is also to be noted that KR and BM BHs are brighter than a \s BH.
\begin{center}
\includegraphics[width=6cm,height=6cm]{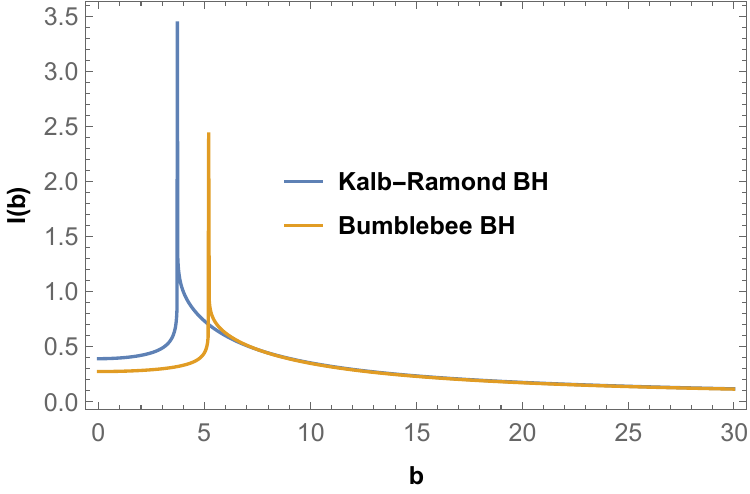}
\parbox[c]{15.0cm}{\footnotesize{\bf Fig~4.} 
A comparison of observed intensity with static accretion. We have taken $\a=\b=0.2$.}
\label{staticcom}
\end{center}
A comparison of observed intensity between KR and BM BHs in Fig. [$4$] reveals that the intensity will always be greater for a KR BH, and as a result, KR BHs will be brighter than BM BHs. Now, we turn our attention towards infalling accretion.
\subsection{Shadow with infalling accretion}
Accretion disks are believed not to be static but have radial velocity, i.e., infalling. Thus, studying shadow with infalling accretion would be closer to reality. The equation for the intensity observed in the case of infalling accretion is the same as that for static accretion with the redshift factor modified to
\be
g_{f}=\frac{k_{\gamma}u_{obs}^{\gamma}}{k_{\zeta}u_{em}^{\zeta}},
\ee
where $k^{\mu}$ represents photon four-velocity, four-velocity of the assymptotic observer is $u_{obs}^{\mu}\equiv(1,0,0,0)$, and accretion four-velocity is $u_{em}^{\mu}$. We know $k_{t}=1/b$ and $k_{\mu}k^{\mu}=0$. This helps us obtain the relation between $k_t$ and $k_r$ as
\begin{equation}
\frac{k_{r}}{k_{t}}=\pm\sqrt{\f{\gr \fr \hr}{\hr-b^2 \fr}}.
\end{equation}
The +(-) sign is when photons move towards(away from) the BH. The four-velocity of accreting matter is
\ba
u_{em}^{t}=\frac{1}{\fr},~~~~~u_{em}^{\theta}=u_{em}^{\phi}=0,~~~~~u_{em}^{r}=-\sqrt{\f{\gr}{\fr}-\gr}.
\ea
\begin{center}
\includegraphics[width=6cm,height=6cm]{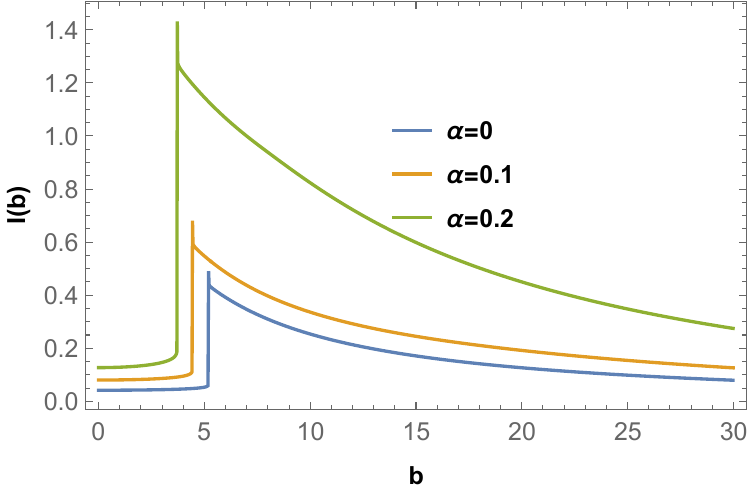}
\includegraphics[width=6cm,height=6cm]{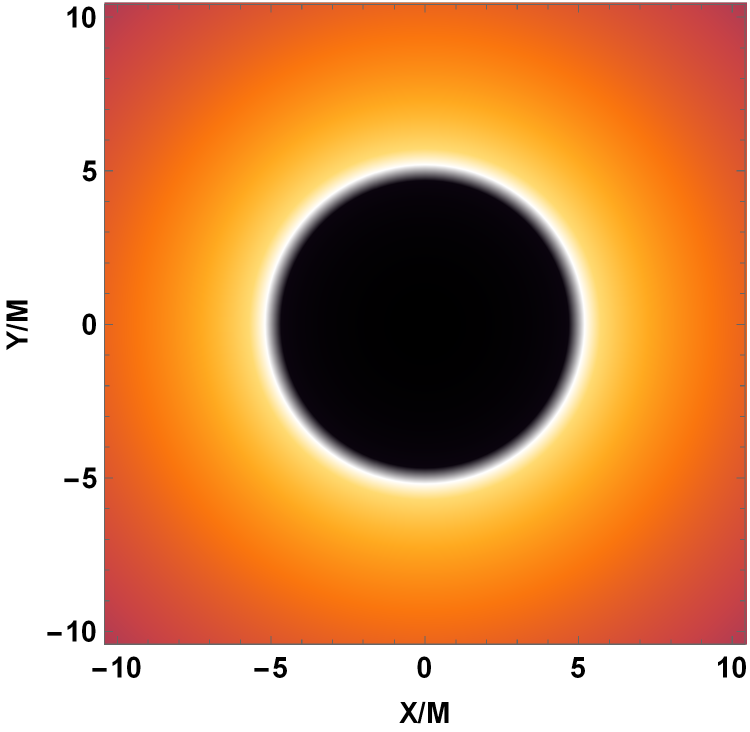}
\includegraphics[width=0.5cm,height=6cm]{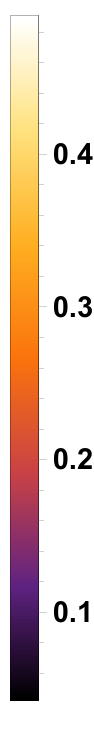}
\includegraphics[width=6cm,height=6cm]{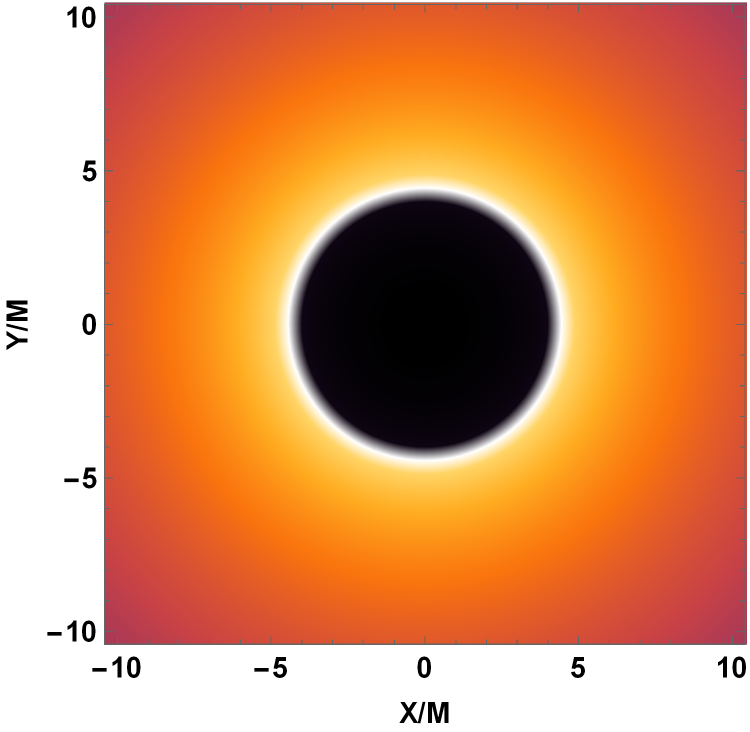}
\includegraphics[width=0.5cm,height=6cm]{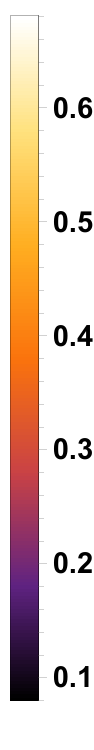}
\includegraphics[width=6cm,height=6cm]{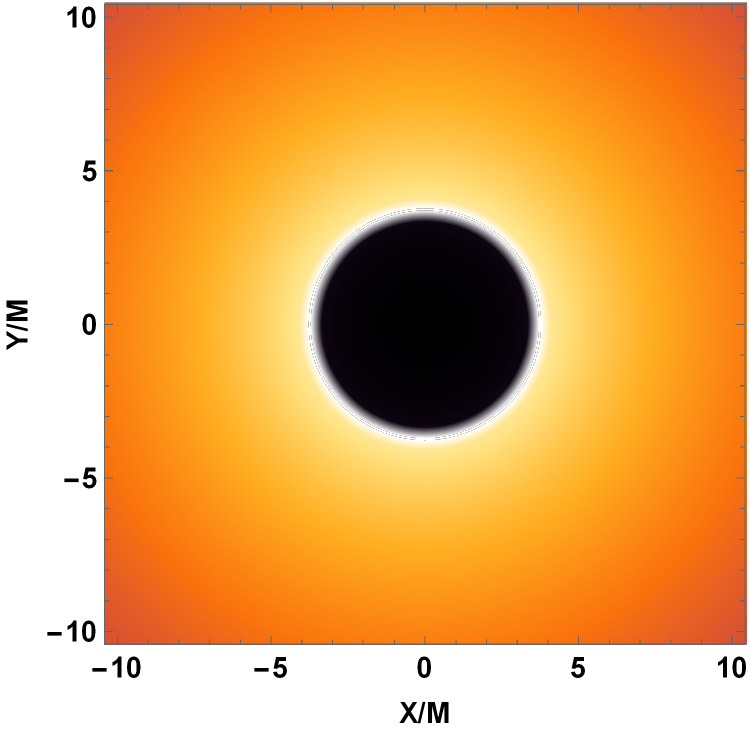}
\includegraphics[width=0.5cm,height=6cm]{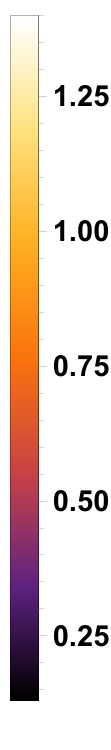}
\parbox[c]{15.0cm}{\footnotesize{\bf Fig~5.} 
Observed intensity and black hole shadows with infalling spherical accretion. The upper right is for $\a=0$, the lower left one is for $\a=0.1$, and the lower right one is for $\a=0.2$.}
\label{infallingkr}
\end{center}
Four-velocities of the photon, distant observer, and accreting matter lead to the following expression of the redshift factor:
\be
g_{f}=\f{\fr \sqrt{\hr}}{\sqrt{\hr}+\sqrt{\hr-b^2 \fr}\sqrt{1-\fr}}.
\ee
Also, the differential proper distance for the infalling case is
\be
dl_{prop}=k_{\mu}u_{em}^{\mu}d\zeta=\frac{k_{t}}{g_{f}|k_{r}|}dr.
\label{properlen}
\ee
With the above expressions at our disposal, we are now in a position to delve into studying the impact of LV on the optical appearance of KR and BM models in Figs. [$5$, $6$]. Similar to the static accretion case, the intensity peaks at the critical impact parameter, albeit the variation with the impact parameter, is smooth compared to the static case. However, the peak value of intensity is much smaller than the static accretion. Although similar to the static accretion, the intensity increases significantly with the KR parameter $\a$, the impact of BM parameter $\b$ is opposite to that in the static case, i.e., increasing $\b$ decreases the intensity. This adverse impact of $\b$ on the intensity is because the radial velocity of photons increases with $\b$, resulting in a decrease in the proper length evident from Eq. [\r{properlen}]. Owing to the Doppler effect, the central region within the photon ring, much darker than the static case, is observed.
\begin{center}
\includegraphics[width=6cm,height=6cm]{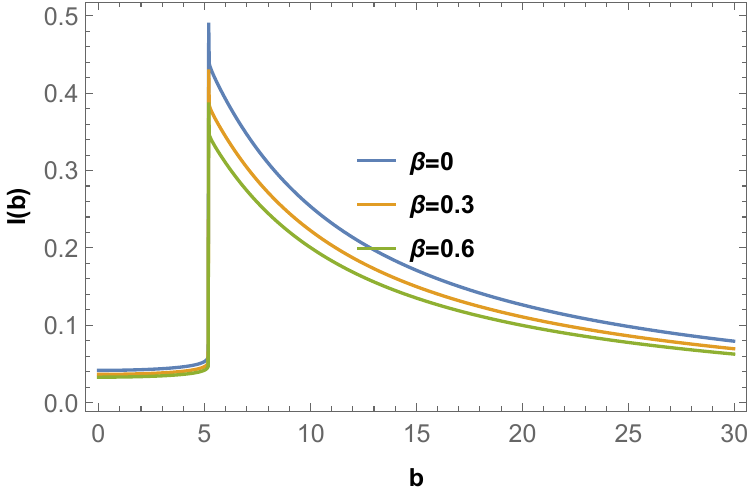}
\includegraphics[width=6cm,height=6cm]{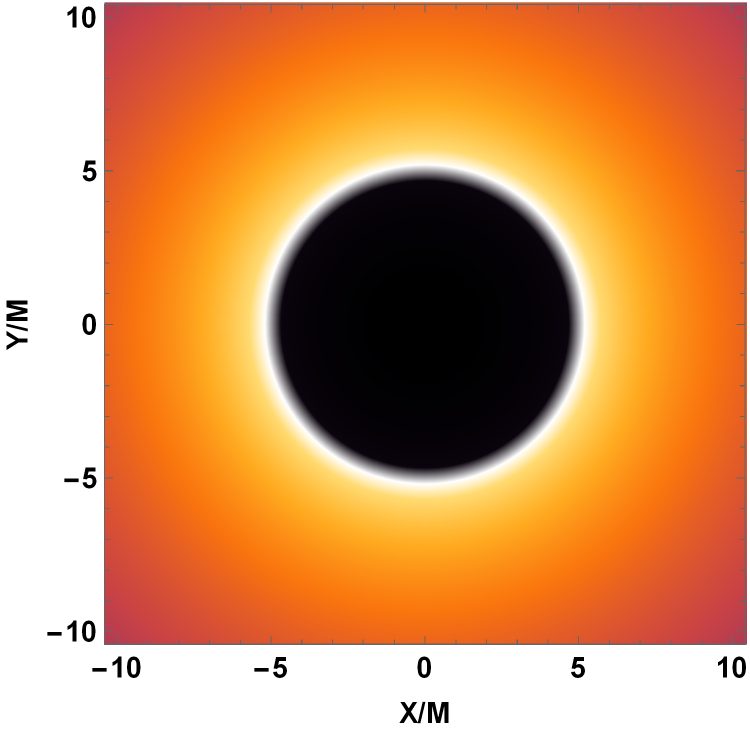}
\includegraphics[width=0.5cm,height=6cm]{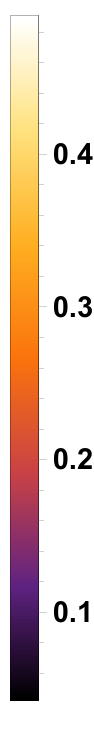}
\includegraphics[width=6cm,height=6cm]{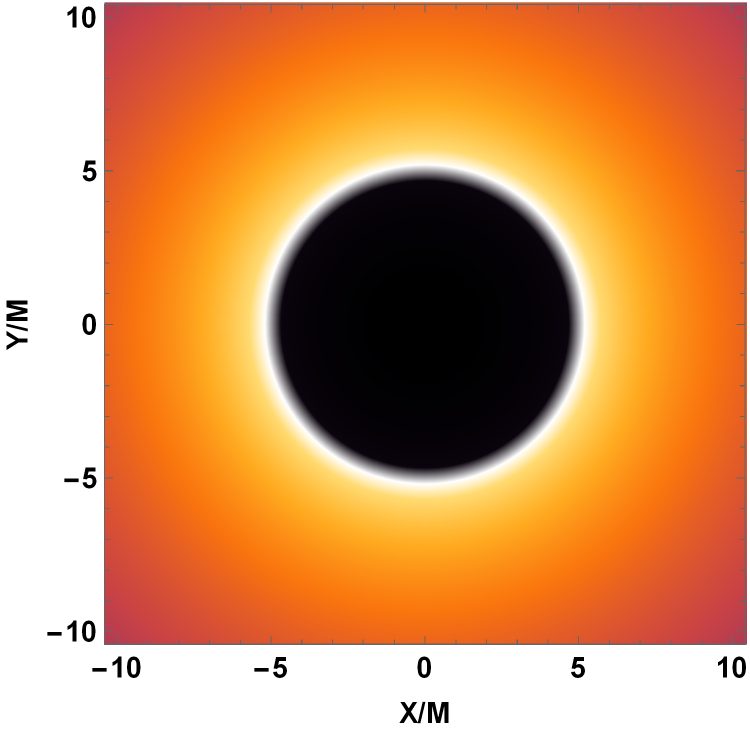}
\includegraphics[width=0.5cm,height=6cm]{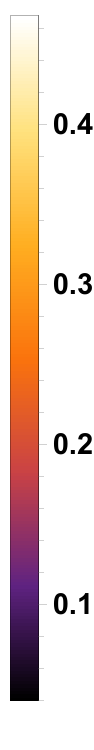}
\includegraphics[width=6cm,height=6cm]{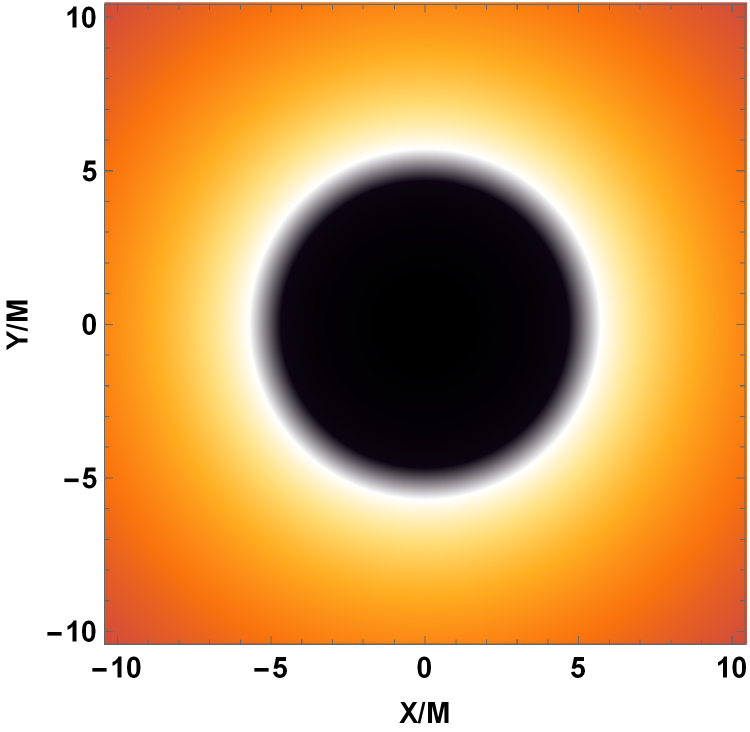}
\includegraphics[width=0.5cm,height=6cm]{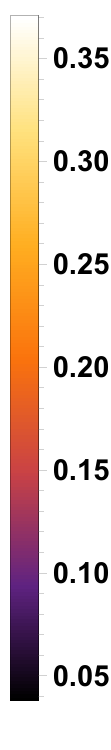}
\parbox[c]{15.0cm}{\footnotesize{\bf Fig~6.} 
Observed intensity and black hole shadows with infalling spherical accretion. The upper right is for $\b=0$, the lower left one is for $\b=0.1$, and the lower right one is for $\b=0.2$.}
\label{infallingbm}
\end{center}
We compare observed intensities for KR and BM BHs. Similar to the static accretion, the intensity for the KR BH is much greater than the BM BH, rendering the BM BH darker than the KR BH or \s BH.
\begin{figure}[H]
\begin{center}
\includegraphics[scale=0.7]{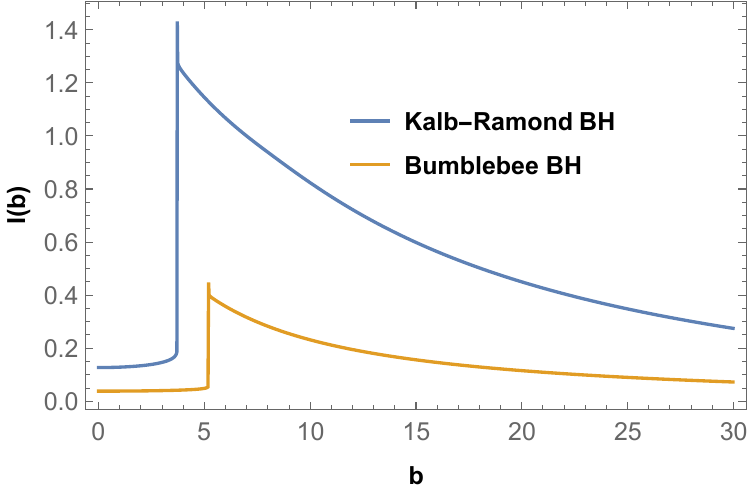}
\caption{A comparison of observed intensity with infalling accretion. We have taken $\a=\b=0.2$. } \label{infallingcom}
\end{center}
\end{figure}
\section{Quasinormal modes}
Quasinormal modes of a BH are complex-valued functions of parameters characterizing the spacetime. As such, it embeds information regarding the central object, making it a possible avenue to probe any deviation from the \s BH. In this section, we explore the impact of LV on QNMs in KR and BM models. Assuming a negligible impact of scalar or electromagnetic perturbation on the background spacetime, we take into account the Klein-Gordon equation for the massless scalar field
\be
\frac{1}{\sqrt{-g}}{\partial \mu}(\sqrt{-g}g^{\mu\nu} \partial_{\nu}\chi) =0,
\label{scalar}
\ee
and the Maxwell equation for the electromagnetic field
\be
\frac{1}{\sqrt{-g}}{\partial \nu }(F_{\rho\sigma}g^{\rho\mu}g^{\sigma\nu}\sqrt{-g})=0,
\label{em}
\ee
where $ F_{\rho\sigma}={\partial \rho}A^\sigma-{\partial \sigma}A^\rho $, $A_\nu$ being the electromagnetic four-potential. We need to reduce Eqs. [\r{scalar}, \r{em}] to a Schr$\ddot{o}$dinger-like equation so that we can employ the $6$th order Pad\'{e} averaged WKB approach. This can be done with the tortoise coordinate defined by $\text{d}r_{\ast}=\frac{\text{d}r}{\fr \gr}$, where $r_*\rightarrow -\infty$ as $r\rightarrow r_h$ and $r_*\rightarrow \infty$ as $r\rightarrow \infty$. This helps us reduce Eqs. [\r{scalar}, \r{em}] to the desired form:
\be
-\frac{\text{d}^2\phi}{\text{d}{r^2_*}}+V(r) \phi=\omega ^{2}\phi.
\label{schrodinger}
\ee
The effective potential has the form
\be
V(r)=|g_{tt}|\left(\frac{\ell(1+\ell)}{r^2}+\frac{1-s^2}{r\sqrt{|g_{tt}|g_{rr}}}
\frac{\text{d}}{\text{d}r}\sqrt{|g_{tt}|g_{rr}^{-1}}\right),
\label{vtotal}
\ee
where $\ell$ and $s$ are the angular momentum and the spin, respectively. $s=0$ gives the effective potential for the scalar perturbation, and the potential for the electromagnetic perturbation is given by $s=1$. With the effective potential, we can obtain quasinormal frequencies with the help of the $6th$ order WKB method by solving the equation \cite{schutz, iyer, iyer1, konoplya1}
\begin{equation}
\frac{\text{i}(\omega ^{2}-V_{0})}{\sqrt{-2V_{0}^{''}}}-\sum ^{6}_{\text{i}=2}\Omega_\text{i}=n+\frac{1}{2},
\label{WKB}
\end{equation}
\begin{figure}[H]
\centering
\subfigure[]{
\label{qnmrefig1}
\includegraphics[width=0.4\columnwidth]{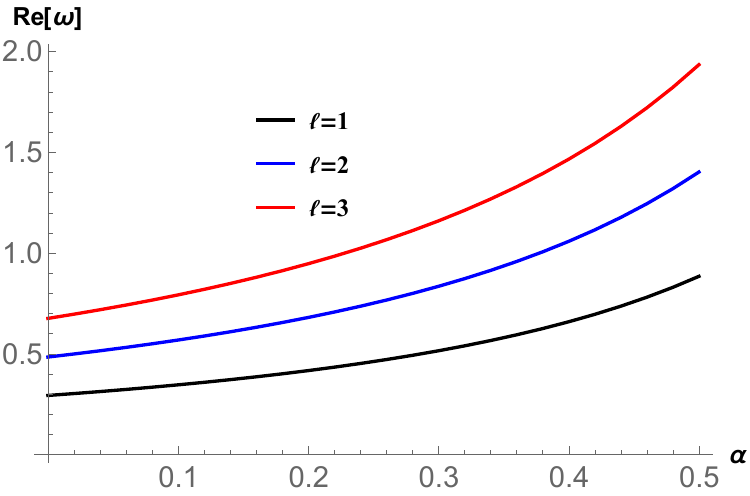}
}
\subfigure[]{
\label{qnmrefig2}
\includegraphics[width=0.4\columnwidth]{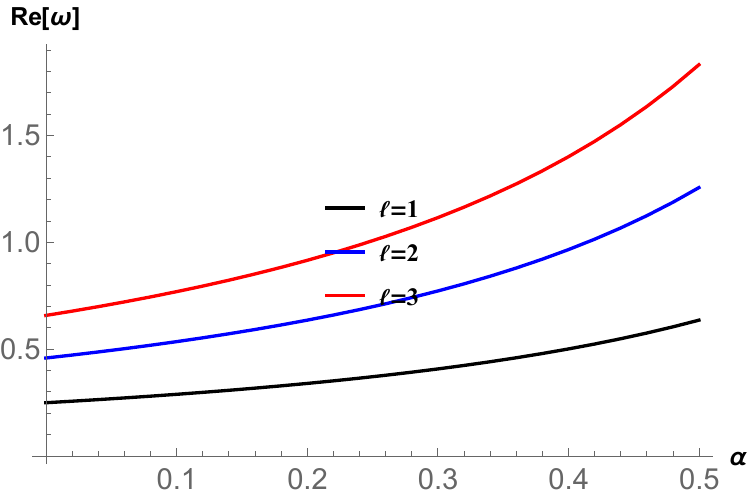}
}
\caption{Variation of gravitational wave frequency with KR parameter $\a$ for different $\ell$.}
\label{rekr}
\end{figure}
\begin{figure}[H]
\centering
\subfigure[]{
\label{qnmrefig1}
\includegraphics[width=0.4\columnwidth]{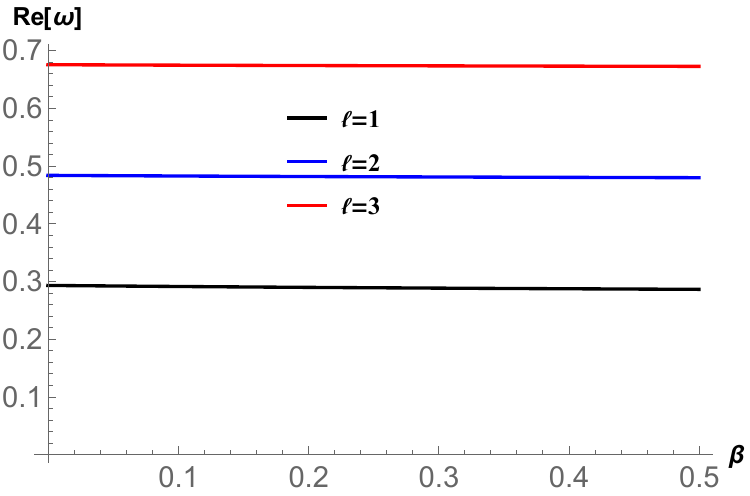}
}
\subfigure[]{
\label{qnmrefig2}
\includegraphics[width=0.4\columnwidth]{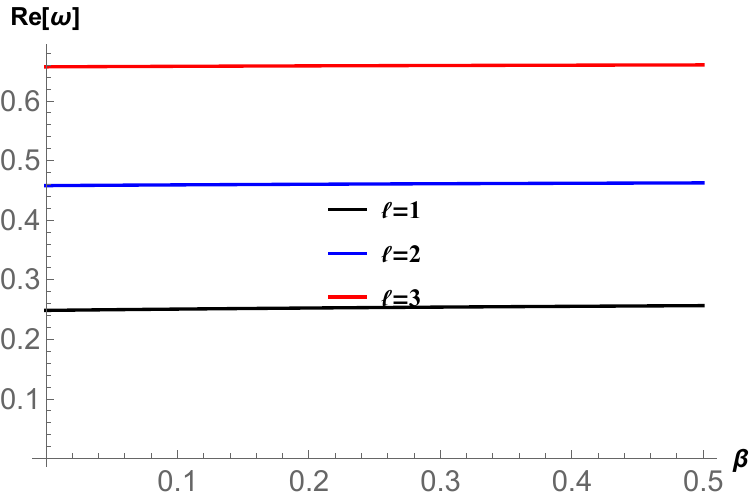}
}
\caption{Variation of gravitational wave frequency with BM parameter $\b$ for different $\ell$.}
\label{rebm}
\end{figure}
\begin{figure}[H]
\centering
\subfigure[]{
\label{qnmrefig1}
\includegraphics[width=0.4\columnwidth]{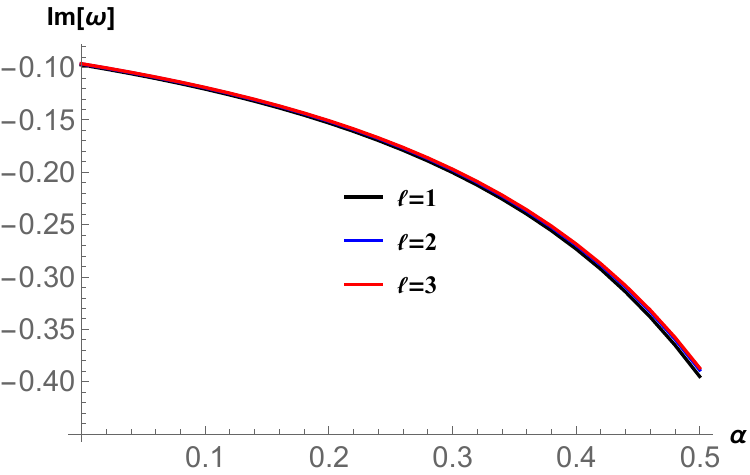}
}
\subfigure[]{
\label{qnmrefig2}
\includegraphics[width=0.4\columnwidth]{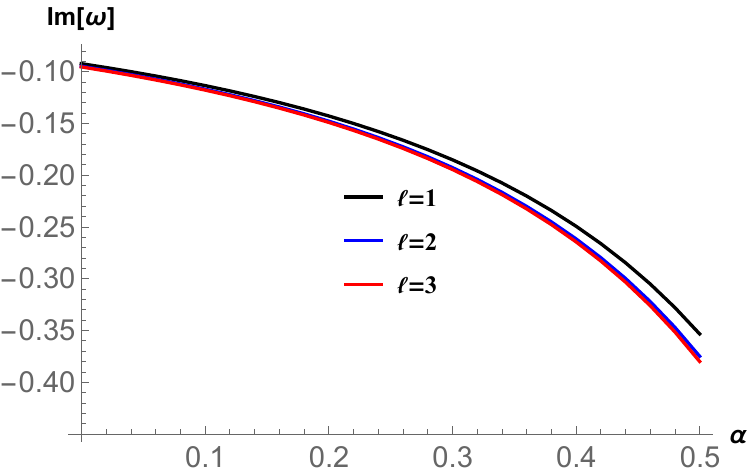}
}
\caption{Variation of decay rate with KR parameter $\a$ for different $\ell$.}
\label{imkr}
\end{figure}
\begin{figure}[H]
\centering
\subfigure[]{
\label{qnmrefig1}
\includegraphics[width=0.4\columnwidth]{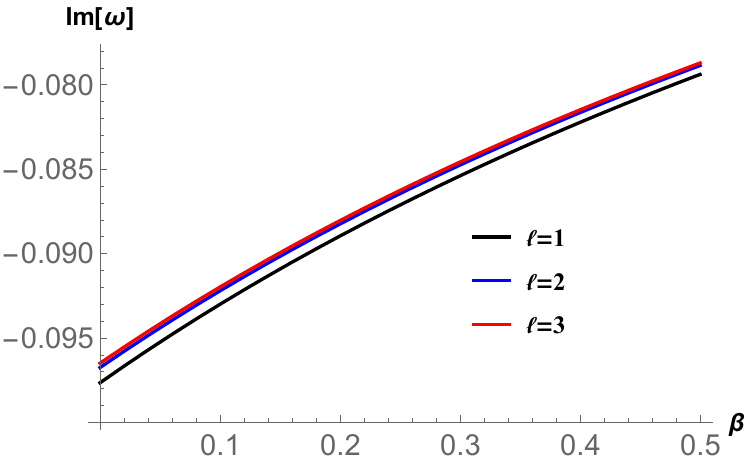}
}
\subfigure[]{
\label{qnmrefig2}
\includegraphics[width=0.4\columnwidth]{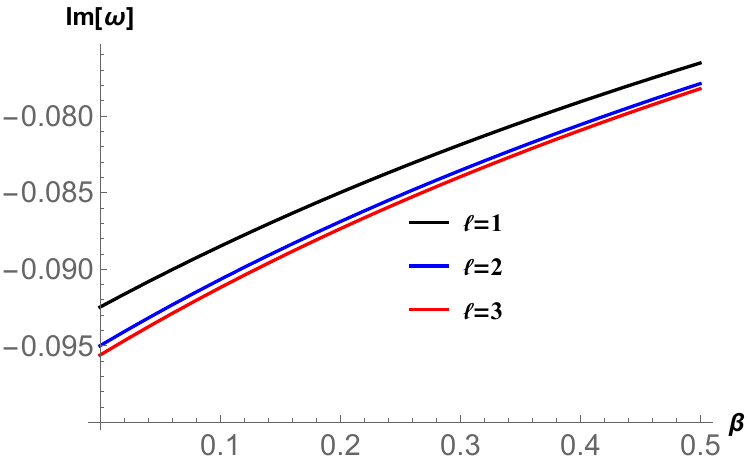}
}
\caption{Variation of decay rate with BM parameter $\b$ for different $\ell$.}
\label{imbm}
\end{figure}
\begin{figure}[H]
\centering
\subfigure[]{
\label{qnmrefig1}
\includegraphics[width=0.4\columnwidth]{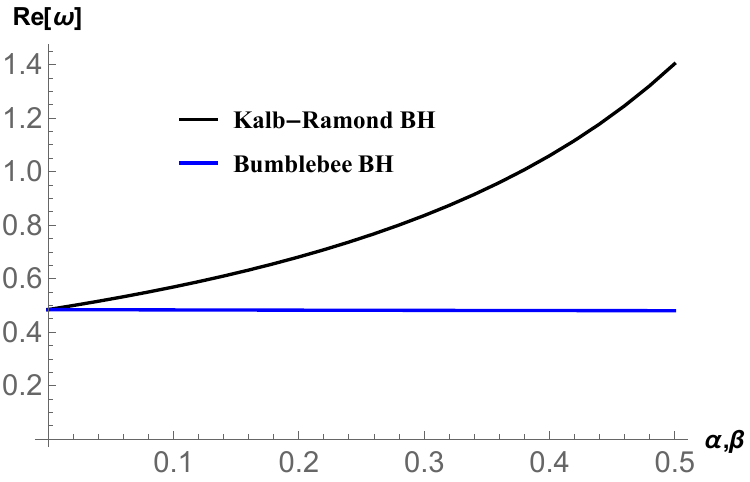}
}
\subfigure[]{
\label{qnmrefig2}
\includegraphics[width=0.4\columnwidth]{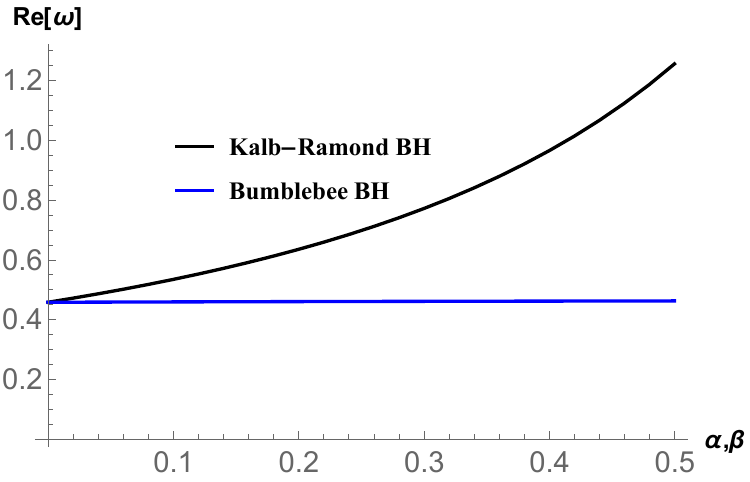}
}
\caption{Comparison of gravitational wave frequency with $\ell=2$.}
\label{recom}
\end{figure}
\begin{figure}[H]
\centering
\subfigure[]{
\label{qnmrefig1}
\includegraphics[width=0.4\columnwidth]{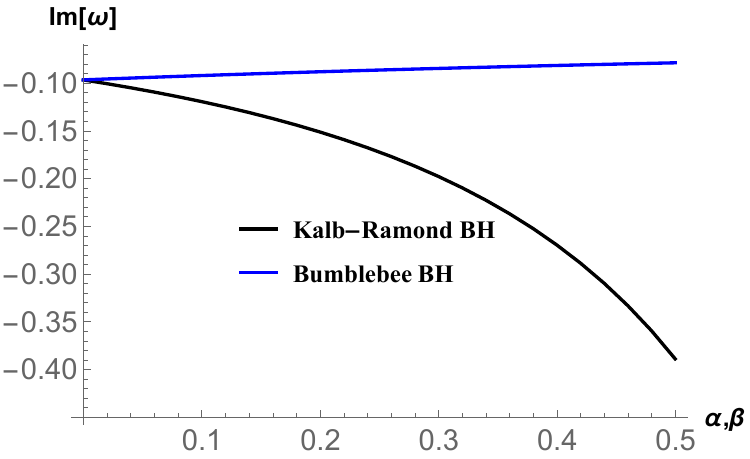}
}
\subfigure[]{
\label{qnmrefig2}
\includegraphics[width=0.4\columnwidth]{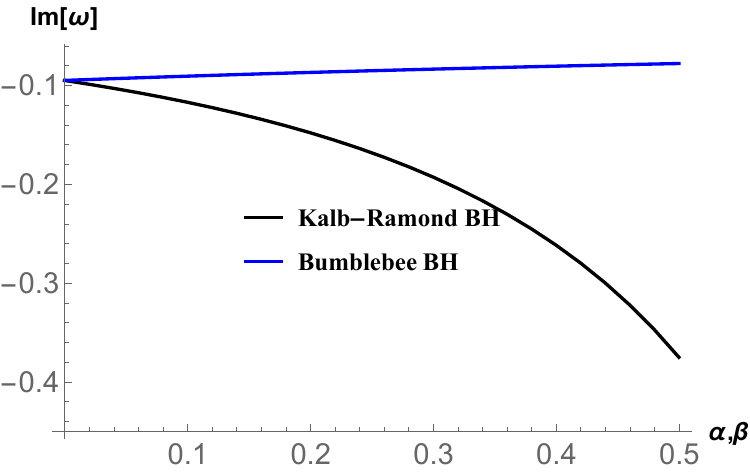}
}
\caption{Comparison of decay rate with $\ell=2$.}
\label{imcom}
\end{figure}
where $V_{0}$ is the peak value of the potential obtained at $r_{*}=r_0$, $V''_{0}=\f{d^2 V(r_*)}{dr_{*}^{2}}|_{r_*=r_0}$, and $\Omega_\text{i}$ are WKB corrections \cite{schutz, iyer, iyer1, konoplya1}. The accuracy of the WKB method can be increased with
Pad\'{e} approximants method \cite{jerzy} where a polynomial of order $k$ in terms of order parameter $\epsilon$ is defined as
\begin{equation}
P_k(\epsilon)=V_0+\Omega_2\epsilon^2+\Omega_4\epsilon^4+\Omega_6\epsilon^6+\ldots - i (n+\frac{1}{2})\sqrt{-2V_{0}^{''}}\left(\epsilon+\Omega_3\epsilon^3+\Omega_5\epsilon^5\ldots\right),
\end{equation}
Putting $\epsilon =1$ in the above equation will lead us to an equation of the form $\omega^2=P_{\tilde{n}/\tilde{m}}(1)$ where \cite{jerzy, konoplya2}
\begin{equation}
P_{\tilde{n}/\tilde{m}}(\epsilon)=\frac{\mathcal{Q}_0+\mathcal{Q}_1\epsilon+\ldots+\mathcal{Q}_{\tilde{n}}\epsilon^{\tilde{n}}}{\mathcal{R}_0+\mathcal{R}_1\epsilon+\ldots+\mathcal{R}_{\tilde{m}}\epsilon^{\tilde{m}}},
\end{equation}
with $\tilde{n}+\tilde{m}=k$. To gauge the impact of LV on the frequency and the decay rate of GW, we plot them against $\a$ and $\b$ for scalar and electromagnetic perturbations. We also tabulate numerical values of QNMs for scalar perturbations in Table [\r{QNMS}] and electromagnetic perturbations in Table [\r{QNMEM}].
\begin{table}[h!]
\begin{adjustwidth}{-1.4cm}{}
\centering
\caption{Quasinormal frequencies for scalar perturbation with $n=0$.}
\setlength{\tabcolsep}{0.5mm}
\begin{tabular}{|c|c|c|c|c|c|c|}
\hline
\text{$\a(\b) $ } & \text{$\ell $=1(KR)} & \text{$\ell $=1(BM)} & \text{$\ell $=2(KR)} & \text{$\ell $=2(BM)} & \text{$\ell $=3(KR)} & \text{$\ell $=3(BM)} \\
\hline
0. & 0.292931\, -0.097661 i & 0.292931\, -0.097661 i & 0.483644\, -0.0967591 i & 0.483644\, -0.0967591 i & 0.675366\, -0.0964997 i & 0.675366\, -0.0964997 i \\
\hline
0.1 & 0.345699\, -0.120733 i & 0.291094\, -0.09301 i & 0.568024\, -0.119524 i & 0.482541\, -0.0922123 i & 0.792124\, -0.119172 i & 0.674579\, -0.0919857 i \\
\hline
0.2 & 0.416377\, -0.153054 i & 0.289557\, -0.0889634 i & 0.680133\, -0.15138 i & 0.481621\, -0.0882512 i & 0.946873\, -0.150884 i & 0.673923\, -0.088051 i \\
\hline
0.3 & 0.514751\, -0.200311 i & 0.288251\, -0.0854008 i & 0.834629\, -0.197898 i & 0.480842\, -0.08476 i & 1.15949\, -0.197169 i & 0.673367\, -0.0845815 i \\
\hline
0.4 & 0.658685\, -0.273324 i & 0.287128\, -0.0822331 i & 1.05789\, -0.269675 i & 0.480173\, -0.0816526 i & 1.46553\, -0.26854 i & 0.67289\, -0.0814921 i \\
\hline
0.5 & 0.884172\, -0.39516 i & 0.286151\, -0.0793927 i & 1.40185\, -0.388943 i & 0.479592\, -0.0788636 i & 1.93457\, -0.387037 i & 0.672476\, -0.0787183 i \\
\hline
\end{tabular}
\label{QNMS}
\end{adjustwidth}
\end{table}

\begin{table}[h!]
\begin{adjustwidth}{-1.45cm}{}
\centering
\caption{Quasinormal frequencies for electromagnetic perturbation with $n=0$.}
\setlength{\tabcolsep}{0.5mm}
\begin{tabular}{|c|c|c|c|c|c|c|}
\hline
\text{$\a(\b) $ } & \text{$\ell $=1(KR)} & \text{$\ell $=1(BM)} & \text{$\ell $=2(KR)} & \text{$\ell $=2(BM)} & \text{$\ell $=3(KR)} & \text{$\ell $=3(BM)} \\
\hline
0. & 0.248251\, -0.0924846 i & 0.248251\, -0.0924847 i & 0.457595\, -0.0950048 i & 0.457595\, -0.0950048 i & 0.656899\, -0.0956163 i & 0.656899\, -0.0956163 i \\
\hline
0.1 & 0.287643\, -0.11364 i & 0.250425\, -0.0885174 i & 0.534146\, -0.117121 i & 0.45885\, -0.0906899 i & 0.768098\, -0.117961 i & 0.657786\, -0.0912194 i \\
\hline
0.2 & 0.338589\, -0.14296 i & 0.252237\, -0.0850152 i & 0.634686\, -0.147964 i & 0.459896\, -0.0869137 i & 0.914633\, -0.149162 i & 0.658526\, -0.0873781 i \\
\hline
0.3 & 0.406394\, -0.185243 i & 0.25377\, -0.0818949 i & 0.771229\, -0.192809 i & 0.460781\, -0.0835727 i & 1.11449\, -0.194602 i & 0.659152\, -0.0839844 i \\
\hline
0.4 & 0.499903\, -0.249383 i & 0.255084\, -0.0790924 i & 0.964788\, -0.261613 i & 0.461539\, -0.0805895 i & 1.39942\, -0.26447 i & 0.659689\, -0.0809576 i \\
\hline
0.5 & 0.63478\, -0.353454 i & 0.256223\, -0.0765576 i & 1.25523\, -0.375054 i & 0.462196\, -0.0779043 i & 1.83038\, -0.380019 i & 0.660154\, -0.0782362 i \\
\hline
\end{tabular}
\label{QNMEM}
\end{adjustwidth}
\end{table}
It is evident from Figs. [\r{rekr}, \r{imkr}] that the KR parameter significantly impacts the frequency as well as the decay rate of GW. The frequency and the decay rate increase with $\a$ for both perturbations. The variation of real and imaginary parts of QNMs with BM parameter $\b$ is shown in Figs. [\r{rebm}, \r{imbm}]. It shows that although the decay rate decreases with $\b$ for both the perturbations, the impact of $\b$ on frequency is not the same for both of them. The frequency decreases with $\b$ for scalar perturbation but increases for electromagnetic perturbation, albeit the variation is marginal. Next, we compare the frequency and the decay rate of KR and BM BHs in Figs. [\r{recom}, \r{imcom}]. It conclusively shows that except at $\a=\b=0$ where we have \s BH, the frequency and decay rate are always greater for a KR BH. Numerical values given in Tables [\r{QNMS}, \r{QNMEM}] are in congruence with conclusions drawn from Figs.[\r{rekr}-\r{imcom}]. \\
\section{time domain profile}
The effect of LV on QNMs can be visualized with a time domain profile. We will be using the integration method described in \cite{gundlach1}, which is based on the finite difference method with the error proportional to the step $\Delta$ used. We can obtain ringdown waveforms with the help of the following equation:
\begin{equation}
\phi_{m,n+1} = -\,\phi_{m, n-1} + \left( \dfrac{\Delta t}{\Delta
\tilde{r}} \right)^2 (\phi_{m+1, n + \phi_{m-1, n}}) + \left(
2-2\left( \dfrac{\Delta t}{\Delta \tilde{r}} \right)^2 - V_m
\Delta t^2 \right) \phi_{m,n},
\end{equation}
with $\phi(\tilde{r}, t) = \phi(m \Delta \tilde{r}, n \Delta t) = \phi_{m,n} $, $V(r(\tilde{r})) = V(\tilde{r},t) = V_{m,n}$. Choosing initial conditions $\phi(r_*,t) = \exp \left[ -\dfrac{(r_*-\hat{r}_{*})^2}{2\sigma^2} \right]$ and $\phi(r_*,t)\vert_{t<0} = 0$
with $r_*=5$, $\hat{r}_*=0.4$ and imposing Von Neumann's stability condition on the ratio $\frac{\Delta t}{\Delta \tilde{r}}$, we obtain time domain profiles of scalar and electromagnetic perturbations.

\begin{figure}[H]
\centering \subfigure[]{ \label{emkr1}
\includegraphics[width=0.45\columnwidth]{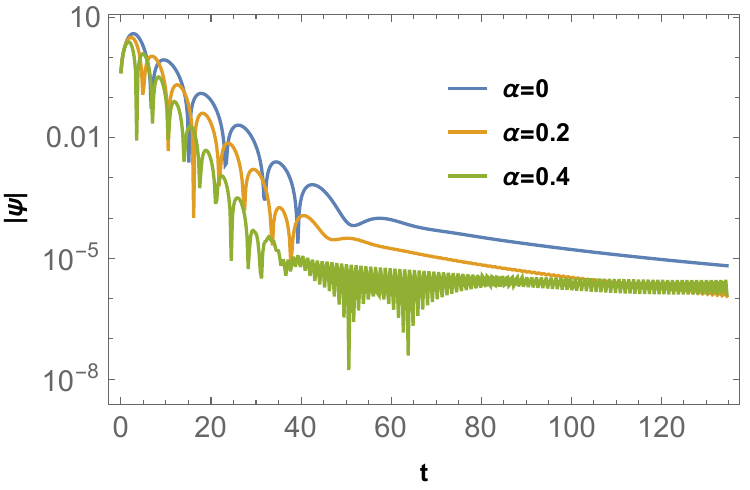}
} \subfigure[]{ \label{emkr2}
\includegraphics[width=0.45\columnwidth]{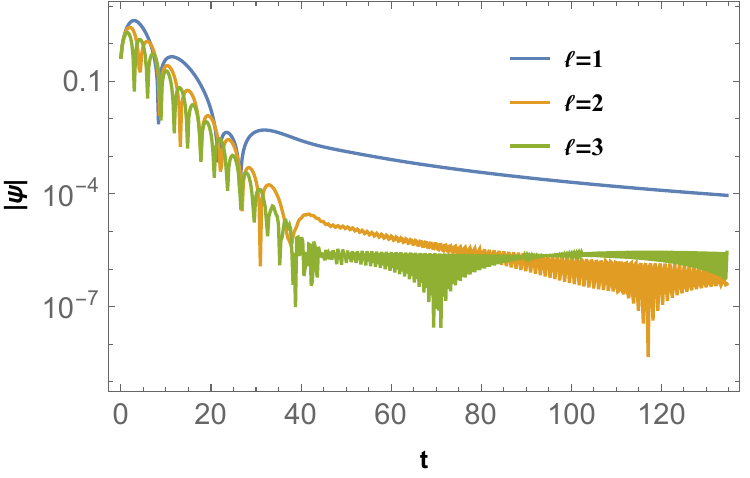}
} \caption{Time domain profile of electromagnetic field for
KR BH. The left one is for $\ell=2$, and the right one is for $\a=0.3$.}
\label{emkr}
\end{figure}

\begin{figure}[H]
\centering \subfigure[]{ \label{scalarkr1}
\includegraphics[width=0.45\columnwidth]{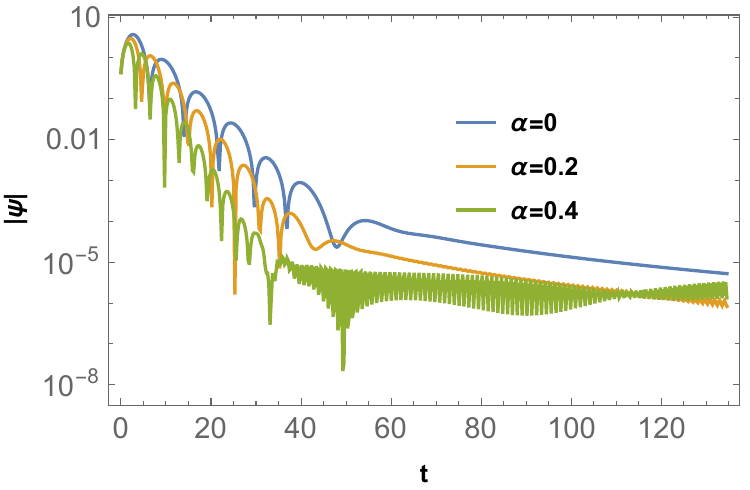}
} \subfigure[]{ \label{scalarkr2}
\includegraphics[width=0.45\columnwidth]{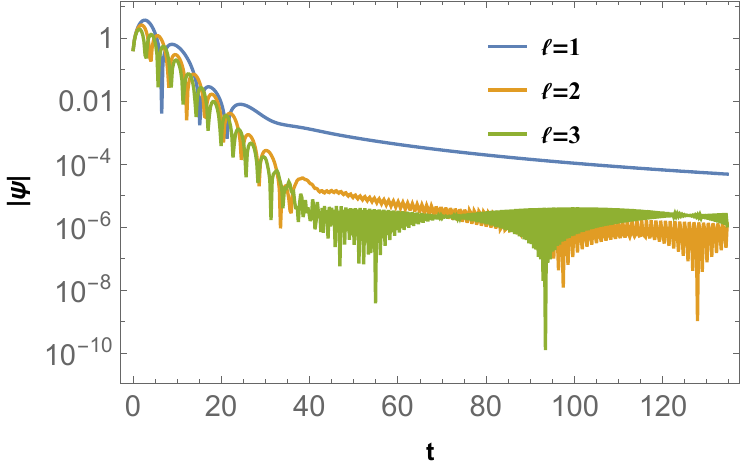}
}
\caption{Time domain profile of scalar field for
KR BH. The left one is for $\ell=2$, and the right one is for $\a=0.3$.}
\label{scalarkr}
\end{figure}

\begin{figure}[H]
\centering \subfigure[]{ \label{embm1}
\includegraphics[width=0.45\columnwidth]{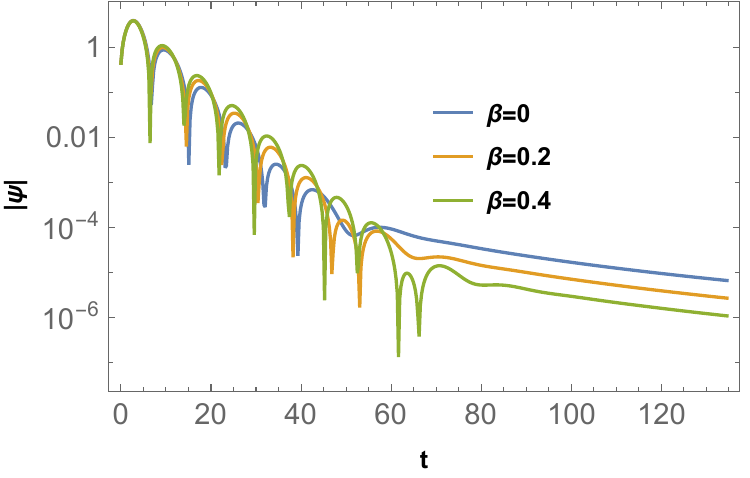}
} \subfigure[]{ \label{embm2}
\includegraphics[width=0.45\columnwidth]{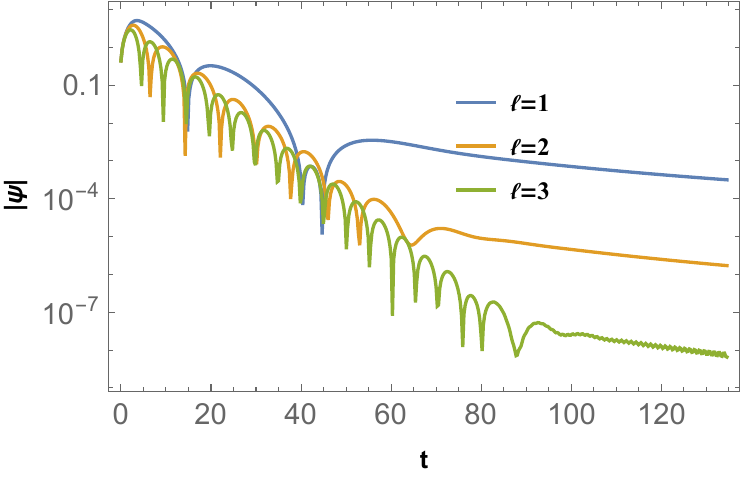}
}
\caption{Time domain profile of electromagnetic field for
BM BH. The left one is for $\ell=2$, and the right one is for $\b=0.3$.}
\label{embm}
\end{figure}

\begin{figure}[H]
\centering \subfigure[]{ \label{scalarkr1}
\includegraphics[width=0.45\columnwidth]{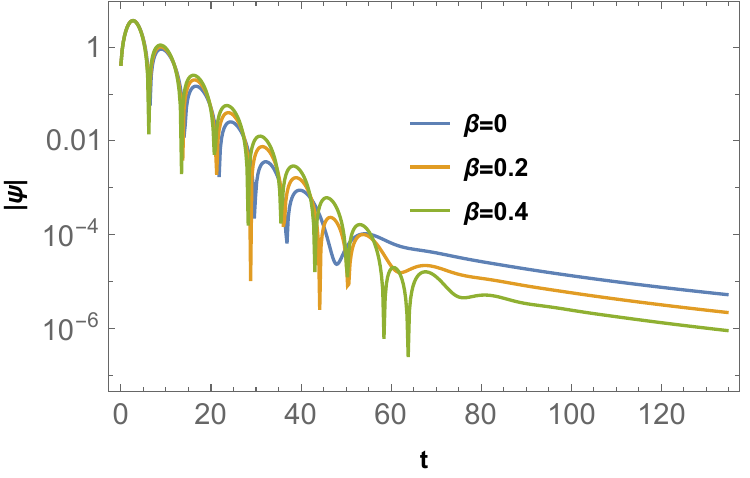}
} \subfigure[]{ \label{scalarkr2}
\includegraphics[width=0.45\columnwidth]{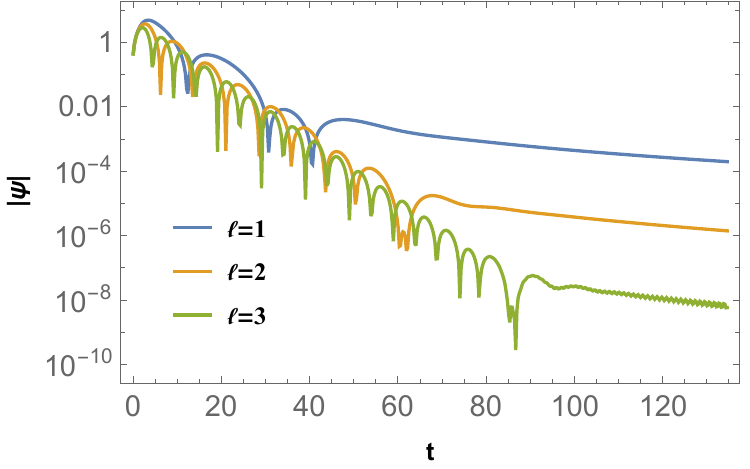}
}
\caption{Time domain profile of scalar field for
BM BH. The left one is for $\ell=2$, and the right one is for $\b=0.3$.}
\label{scalarbm}
\end{figure}

\begin{figure}[H]
\centering \subfigure[]{ \label{scalarcom}
\includegraphics[width=0.45\columnwidth]{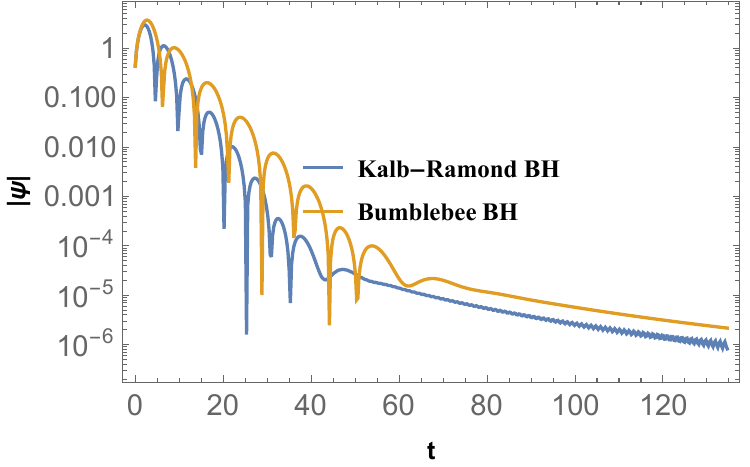}
} \subfigure[]{ \label{emcom}
\includegraphics[width=0.45\columnwidth]{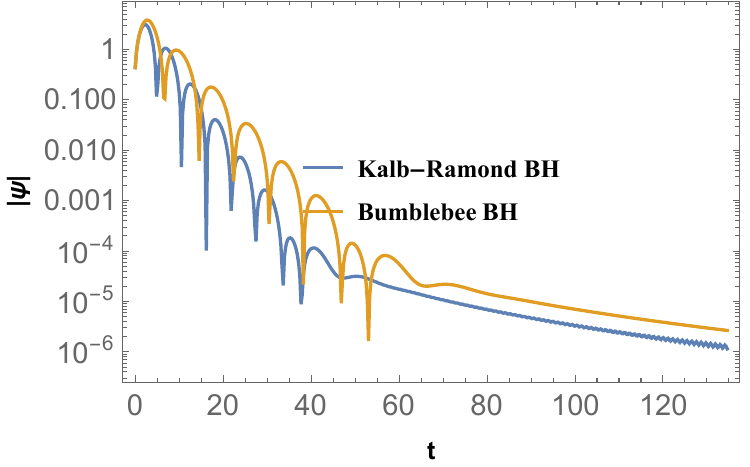}
}
\caption{Comparison of time domain profiles with $\a=\b=0.2$ and $\ell=2$.}
\label{com}
\end{figure}
Similar to the previous section, we observe from Figs. [\r{emkr1}, \r{scalarkr1}] that the frequency and the decay rate increases with $\a$ for both perurbations. However, it is evident from Figs. [\\r{emkr2}, \r{scalarkr2}] that the decay rate increases with $\ell$ for electromagnetic perturbation but decreases for scalar field. Figs. [\r{embm}, \r{scalarbm}] reinforce our findings in the previous section regarding the effect of the BM parameter on QNMs. A comparison of time-domain profiles between KR and BM BHs is shown in Fig. [\r{com}]. It can be conclusively inferred from the figure and Table [\r{QNMS}, \r{QNMEM}] that GWs originated from a KR BH due to scalar or electromagnetic perturbation have larger frequency and decay faster than GWs originating from a BM BH or a \s BH.
\section{Greybody factor and Hawking spectrum}
We will explore LV's possible impacts on the transmission coefficient or GF and Hawking spectrum. BHs emit radiation \c{HAWKING} and possess a temperature \c{BEK, KEIF}. The temperature associated with the BH is
\begin{equation}
T_H=\frac{1}{4\pi \sqrt{-g_{tt}g_{rr}}}\frac{dg_{tt}}{dr}|_{r=r_h}.
\end{equation}
We have $g_{tt}=-\fr$ and $g_{rr}=\f{1}{\gr}$. The Hawking temperature of a KR BH is $T_H=\f{1}{8\pi M \lt(1-\a \rt)^2}$ and for a BM BH $T_H=\f{1}{8\pi M \sqrt{1+\b}}$. It is evident from these expressions that LV in the KR model increases the Hawking temperature but impacts adversely in the case of the BM model. The radiation observed by a local observer differs from that observed by a distant one, owing to the redshift factor. The radiation received by an asymptotic observer is described by greybody distribution. There exists a number of methods to calculate GF \cite{qn32,qn36,qn37,qn38,qn39,qn40,qn41,qn42,qn43,qn44,qn45,qn46}. Visser and Boonserm gave an elegant method in \c{GB, GB1, GB2} to lower bound GF as
\be
T\left( \omega\right) \geq \sec h^{2}\left( \frac{1}{2w}\int_{-\infty}^{+\infty }Vdr_{\ast }\right).
\ee
The above equation can also be written as
\be
T\left( \omega\right) \geq \sec h^{2}\left( \frac{1}{2w}%
\int_{r_{h}}^{+\infty }V\f{dr}{\sqrt{\fr \gr}}\right).
\ee
With the help of the above equation, the lower bound of GF for a KR BH is obtained as $T\left( \omega\right)=\sec h^2\lt[-\frac{2 (\alpha -1) \ell (\ell+1)+s^2-1}{8 (\alpha -1)^2 M \omega }\rt]$ and for a BM BH, the lower bound is $T\left( \omega\right)=\sec h^2\lt[\frac{2 \sqrt{\beta +1} \ell (\ell+1)-\sqrt{\frac{1}{\beta +1}} \left(s^2-1\right)}{8 M \omega }\rt]$. We want to probe the impact of LV on GF and compare its effect in KR and BM models. The qualitative nature of variation of GF with KR parameter $\a$ is shown in Fig. [\r{greykr}], and for BM BH, it is shown in Fig. [\r{greybm}]. A comparison between KR and BM BHs is illustrated in Fig. [\r{greycom}].

\begin{figure}[H]
\centering \subfigure[]{ \label{greykr1}
\includegraphics[width=0.45\columnwidth]{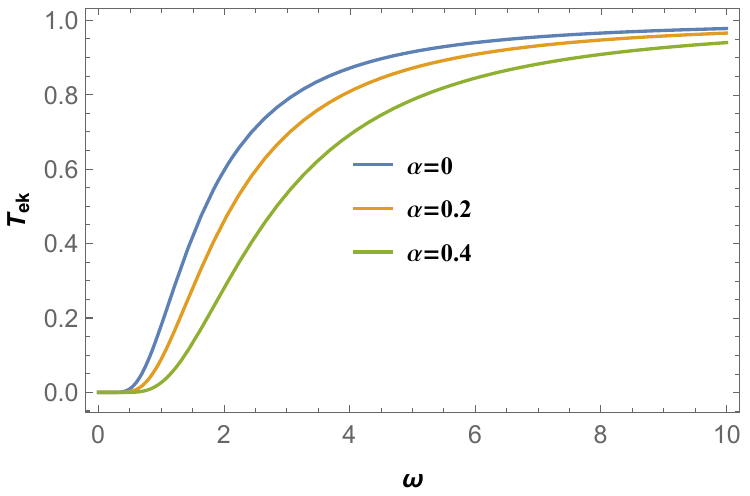}
} \subfigure[]{ \label{greykr2}
\includegraphics[width=0.45\columnwidth]{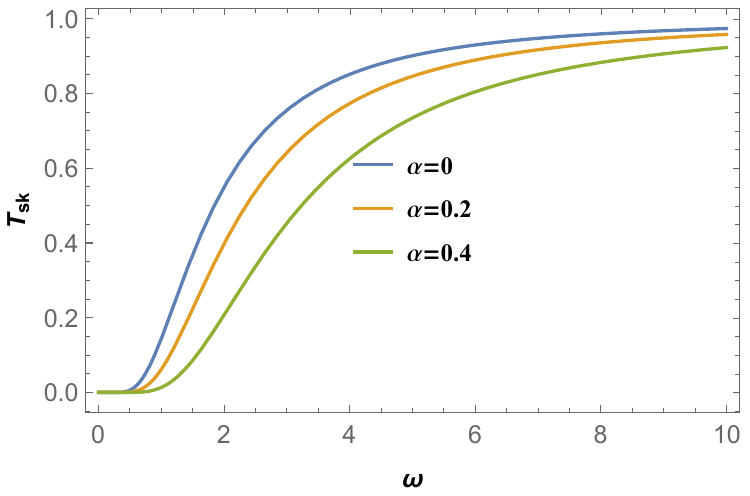}
}
\caption{Greybody bounds for KR BH. The left one is for the electromagnetic field, and the right one is for the scalar field. We have taken $\ell=2$.}
\label{greykr}
\end{figure}

\begin{figure}[H]
\centering \subfigure[]{ \label{greykr1}
\includegraphics[width=0.45\columnwidth]{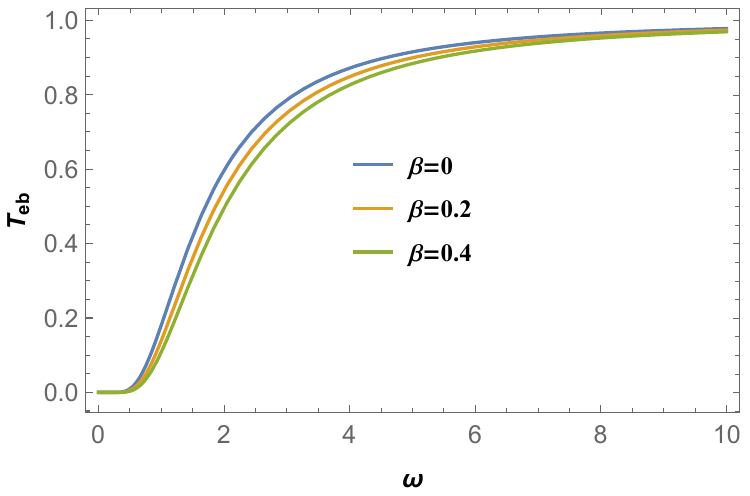}
} \subfigure[]{ \label{greykr2}
\includegraphics[width=0.45\columnwidth]{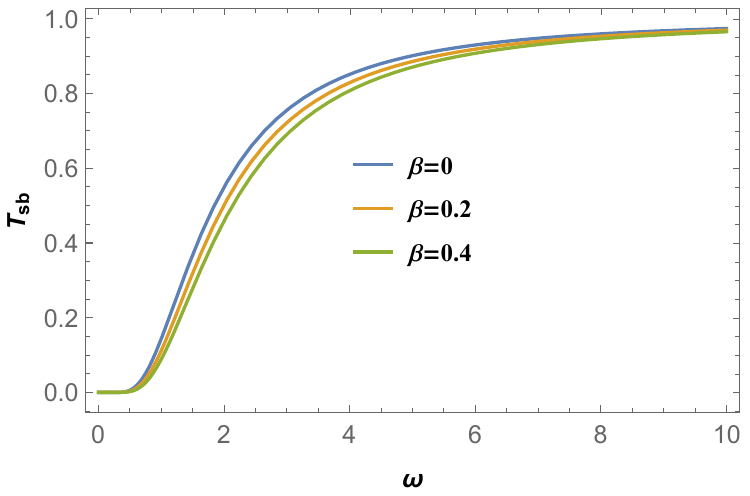}
}
\caption{Greybody bounds for BM BH. The left one is for the electromagnetic field, and the right one is for the scalar field. We have taken $\ell=2$.}
\label{greybm}
\end{figure}

\begin{figure}[H]
\centering \subfigure[]{ \label{greykr1}
\includegraphics[width=0.45\columnwidth]{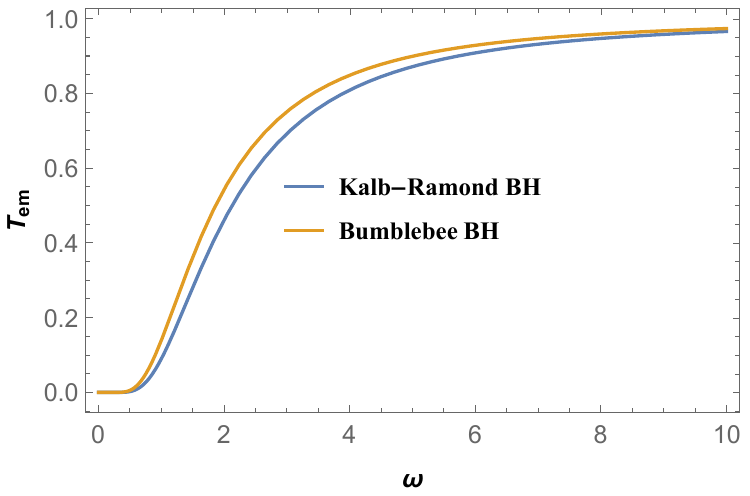}
} \subfigure[]{ \label{greykr2}
\includegraphics[width=0.45\columnwidth]{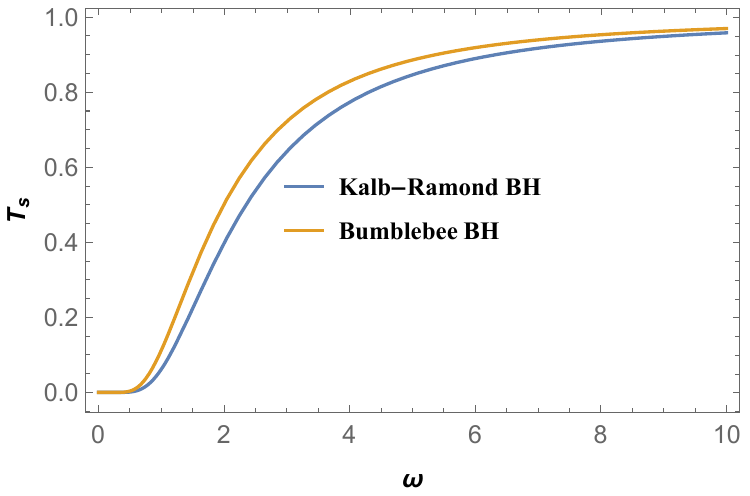}
}
\caption{Comparison of Greybody bounds. The left one is for the electromagnetic field, and the right one is for the scalar field. We have taken $\ell=2$ and $\a=\b=0.2$.}
\label{greycom}
\end{figure}
It can be inferred from Figs. [\r{greykr}, \r{greybm}] that the transmission probability decreases with $\a$ and $\b$. However, the variation of GF with $\b$ is smaller. Our comparison of GFs between two BHs shows that the probability of thermal radiation reaching spatial infinity is less for a KR BH. Next, we investigate the power emitted in the $\ell$th mode which is given by \c{yg2017, fg2016}
\begin{equation}\label{pl}
P_\ell\left(\omega\right)=\frac{A}{8\pi^2}T(\omega)\frac{\omega^3}{e^{\omega/T_{H}}-1}.
\end{equation}
Here, $A$ is the horizon area \cite{yg2017}.
\begin{figure}[H]
\centering \subfigure[]{ \label{greykr1}
\includegraphics[width=0.45\columnwidth]{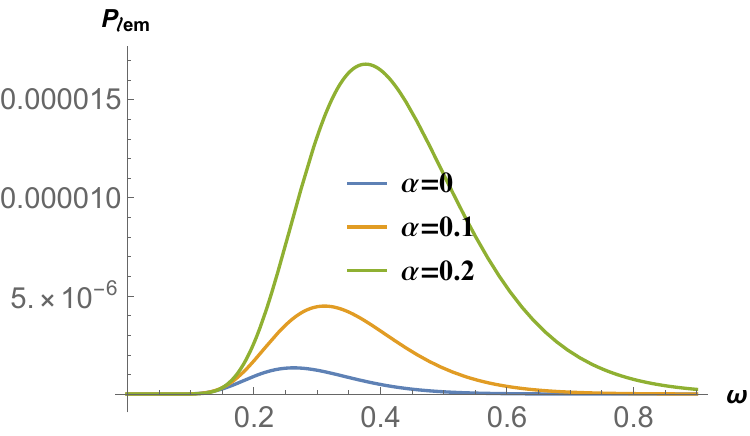}
} \subfigure[]{ \label{greykr2}
\includegraphics[width=0.45\columnwidth]{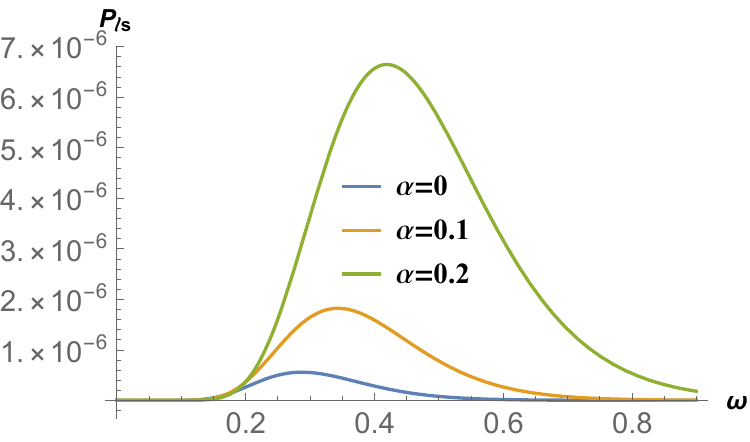}
}
\caption{Variation of power emitted by KR BH in the $\ell$th mode with $\omega$ for different values of $\a$. The left one is for the electromagnetic field, and the right one is for the scalar field. We have taken $\ell=1$.}
\label{plkr}
\end{figure}

\begin{figure}[H]
\centering \subfigure[]{ \label{greykr1}
\includegraphics[width=0.45\columnwidth]{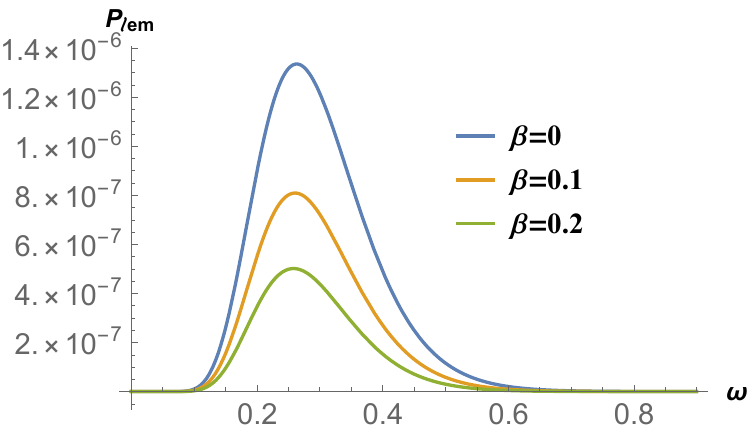}
} \subfigure[]{ \label{greykr2}
\includegraphics[width=0.45\columnwidth]{pl_em_bm_l_1.pdf}
}
\caption{Variation of power emitted by BM BH in the $\ell$th mode with $\omega$ for different values of $\b$. The left one is for the electromagnetic field, and the right one is for the scalar field. We have taken $\ell=1$.}
\label{plbm}
\end{figure}

\begin{figure}[H]
\centering \subfigure[]{ \label{greykr1}
\includegraphics[width=0.45\columnwidth]{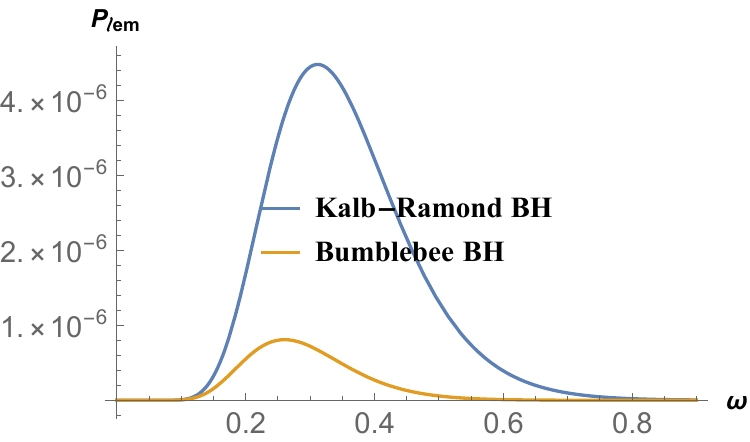}
} \subfigure[]{ \label{greykr2}
\includegraphics[width=0.45\columnwidth]{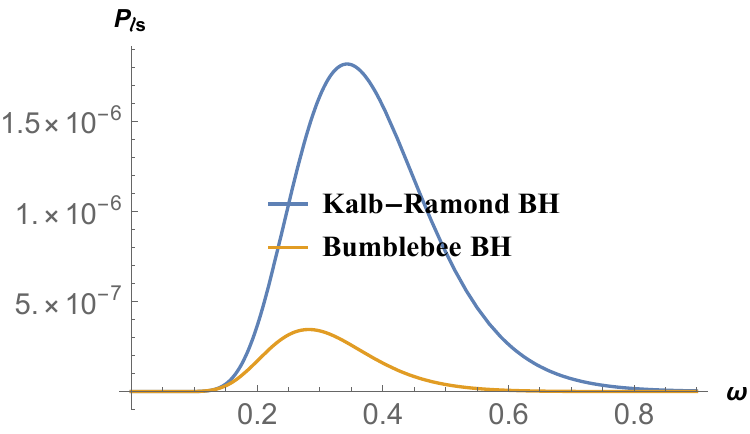}
}
\caption{Comparison of power emitted in $\ell$th mode. The left one is for the electromagnetic field, and the right one is for the scalar field. We have taken $\ell=2$ and $\a=\b=0.2$.}
\label{plcom}
\end{figure}
A qualitative study of the variation of power emitted with $\omega$ for different values of LV parameters $\a$ and $\b$ as shown in Figs. [\r{plkr}, \r{plbm}] reveal that the power emitted by a KR BH increases with the LV parameter and the position of the peak shifts towards the right. On the other hand, the energy emitted decreases with the LV parameter for a BM BH, and the frequency where the power peaks shifts towards the left. A comparison between KR and BM BHs regarding the thermal power they emit reveals that an observer at spatial infinity will receive much more thermal radiation from a KR BH than a BM BH.
\section{weak gravitational lensing}
\begin{figure}[H]
\begin{center}
\includegraphics[scale=0.7]{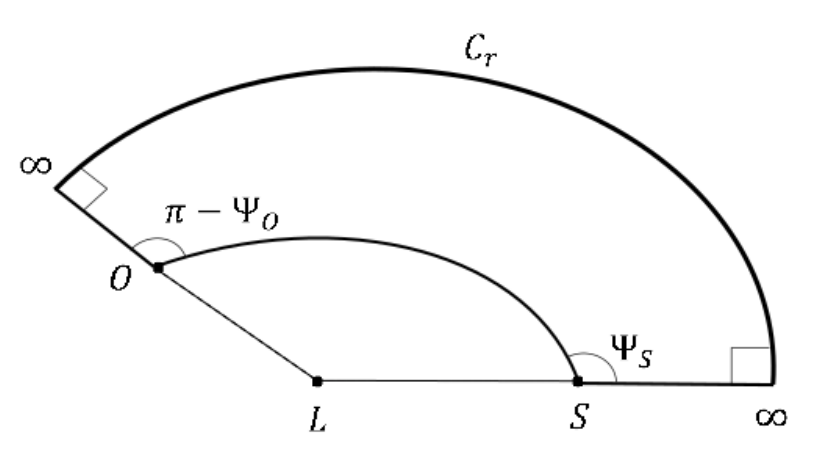}
\caption{Schematic diagram of the quadilateral ${}_O^{\infty}\Box_{S}^{\infty}$. }\label{lensing}
\end{center}
\end{figure}

BHs, acting as lenses, force even light rays to deviate from their straight-line path. Due to its dependence on the underlying spacetime, gravitational lensing becomes a potent tool to gauge the impact of LV parameters. We will study weak gravitational lensing in the background of KR and BM BHs with the help of the Gauss-Bonnet theorem proposed in \c{GW}. The deflection angle can be obtained using the formula \c{ISHIHARA1, CARMO}
\be
\gamma_D=-\int\int_{{}_R^{\infty}\Box_{S}^{\infty}} K
dS,\label{deflectionangle}
\ee
where ${}_O^{\infty}\Box_{S}^{\infty}$ is the quadrilateral shown in Fig. (\ref{lensing}) and K is the Gaussian curvature. We consider null geodesics where $ds^2=0$ resulting in
\begin{equation}
dt^2= \zeta_{ij}dx^i dx^j,\label{metric2}
\end{equation}
where
\begin{equation}
\zeta_{ij}dx^i dx^j=\frac{1}{\fr \gr}dr^2+\frac{\hr}{\fr}\left(d\theta^2+\sin^2\theta\,
d\phi^2\right). \label{metric3}
\end{equation}
We can obtain the Gaussian curvature using the expression \cite{WERNER}
\begin{eqnarray}\nonumber
K&=&\frac{{}^{3}R_{r\phi r\phi}}{\zeta},\\\nonumber
&=&\frac{1}{\sqrt{\zeta}}
\left(\frac{\partial}{\partial \phi}
\left(\frac{\sqrt{\zeta}}{\zeta_{rr}}{}^{(3)}\Gamma^{\phi}_{rr}\right)
- \frac{\partial}{\partial r}\left(\frac{\sqrt{\zeta}}{\zeta_{rr}}{}^{(3)}
\Gamma^{\phi}_{r\phi}\right)\right)\\
,\label{gaussian}
\end{eqnarray}
where $\zeta=\f{\hr}{\fr^2 \gr}$, $\zeta_{rr}=\f{1}{\fr \gr}$, ${}^{(3)}\Gamma^{\phi}_{rr}=0$, and ${}^{(3)}\Gamma^{\phi}_{r\phi}=\f{1}{2} \f{h'(r)\fr-\hr f'(r)}{\fr \hr}$, $'$ representing derivative with respect to r. The final expression for the Gaussian curvature becomes
\be
K=-\f{\fr \sqrt{gr}}{\sqrt{\hr}} \f{\pt}{\pt r}\lt(\sqrt{\fr \gr}\f{\pt}{\pt r}\sqrt{\f{\hr}{\fr}}\rt).
\ee
The above expression lets us compute the Gaussian curvature for the BHs under consideration. The Gaussian curvature for a KR BH is $K=\frac{M (3 M-2 (\alpha +1) r)}{r^4}+\mathcal{O}\left( \f{\a^2 M}{r^3}\right)$ and for a BM BH, it is $K=\frac{(\beta -1) M (2 r-3 M)}{r^4}+\mathcal{O}\left( \f{\a^2 M}{r^3}\right)$. Assuming $r_0$ being the closest approach to the BH, we rewrite Eq. (\ref{deflectionangle}) as \c{ONO1}
\be
\int\int_{{}_O^{\infty}\Box_{S}^{\infty}} K dS =
\int_{\phi_S}^{\phi_O}\int_{\infty}^{r_0} K \sqrt{\zeta}dr
d\phi,\label{Gaussian}
\ee
We intend to obtain higher-order corrections in the deviation angle. To this end, we first obtain the deflection angle assuming a straight-line trajectory where $r=\frac{b}{sin\phi}$ and then using that deflection angle and the trajectory given in \cite{CRISNEJO} as
\be
u=\f{1}{r}=\frac{\sin\phi}{b} + \frac{M(1-\cos\phi)^2}{b^2}-\frac{M^2(60\phi\,
\cos\phi+3\sin3\phi-5\sin\phi)}{16b^3}+\mathcal{O}\left( \frac{M^2\alpha}{b^5}\right),
,\label{uorbit}
\ee
we obtain higher-order corrections. The initial deflection angle for a KR BH is $\gamma_D^0=M^2 \left(\frac{3 \pi }{4 b^2}-\frac{9 \pi \alpha }{8 b^2}\right)+M \left(\frac{4}{b}-\frac{2 \alpha }{b}\right)+\mathcal{O}\left( \frac{M^3}{b^3},\frac{M^3\a}{b^3}\right)$ and for a BM BH is $\gamma_D^0=M^2 \left(\frac{3 \pi }{4 b^2}-\frac{3 \pi \alpha }{8 b^2}\right)+M \left(\frac{4}{b}-\frac{2 \alpha }{b}\right)+\mathcal{O}\left( \frac{M^3}{b^3},\frac{M^3\b}{b^3}\right)$. Now, integrating $\phi$ from $0$ to $\pi+\gamma_D^0$ and using the trajectory [\r{uorbit}], we obtain the deflection angle for a KR BH as
\be
\gamma_D=\frac{128 M^3}{3 b^3}+\frac{15 \pi M^2}{4 b^2}+\alpha \left(-\frac{104 M^3}{3 b^3}-\frac{21 \pi M^2}{8 b^2}-\frac{2 M}{b}\right)+\frac{4 M}{b}+\mathcal{O}\left( \frac{M^4}{b^4},\frac{M^4\a}{b^4}\right),
\label{defkr}
\ee
and the deflection angle for a BM BH is
\be
\gamma_D=\frac{128 M^3}{3 b^3}+\frac{15 \pi M^2}{4 b^2}+\beta \left(-\frac{64 M^3}{3 b^3}-\frac{15 \pi M^2}{8 b^2}-\frac{2 M}{b}\right)+\frac{4 M}{b}+\mathcal{O}\left( \frac{M^4}{b^4},\frac{M^4\b}{b^4}\right).
\label{defbm}
\ee
One can observe from analytical expressions in Eqs. [\r{defkr}, \r{defbm}] that the impact of LV in KR and BM models on the gravitational lensing is the same, i.e., the deflection angle decreases with $\a$ and $\b$.
\begin{figure}[H]
\centering \subfigure[]{ \label{defkr}
\includegraphics[width=0.4\columnwidth]{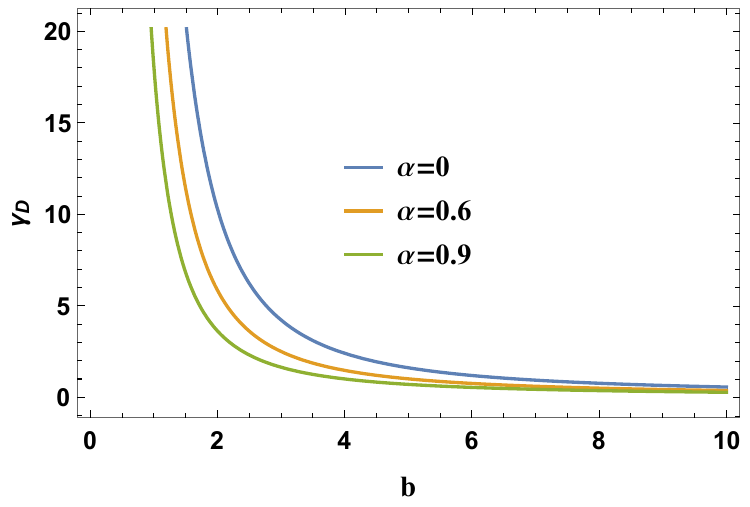}
} \subfigure[]{ \label{defbm}
\includegraphics[width=0.4\columnwidth]{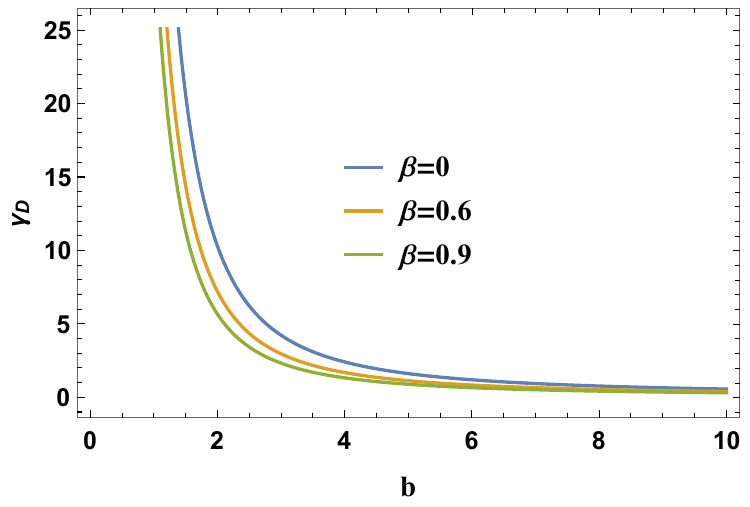}
}\subfigure[]{ \label{defcom}
\includegraphics[width=0.4\columnwidth]{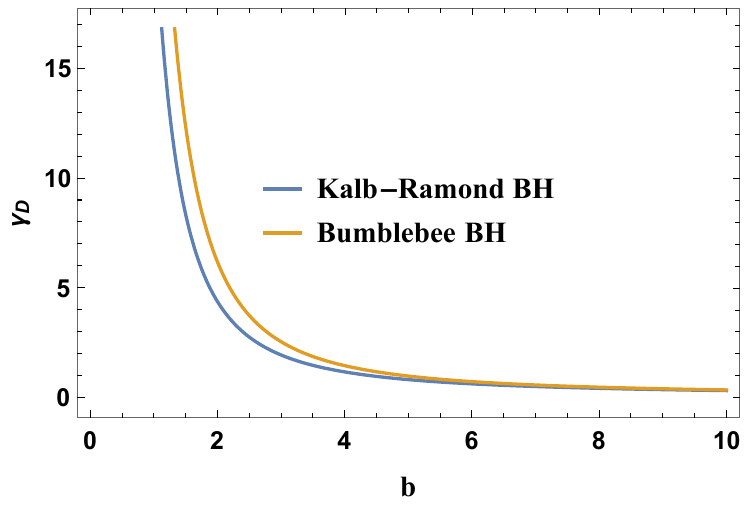}
}
\caption{Deflection angle for weak gravitational lensing. The left
panel is for KR BH, the middle one is for BM BH, and the right one illustrates the comparison.}\label{weak}
\end{figure}
The qualitative nature of variation of deflection angle with the impact parameter $b$ for different values of KR parameter is illustrated in Fig. [\r{defkr}], for BM BH it is shown in Fig. [\r{defbm}], and a comparison between KR and BM BHs is carried out in Fig. [\r{defcom}]. Although the deflection angle decreases for both types of BHs, the variation is less for a BM BH. The comparison between KR and BM BHs reveals that a light ray deviates more from its straight-line path when passing by a BM BH. However, the deviation for both BHs under consideration is less than the \s BH.
\section{Conclusions}
Probing LV through its impact on various astrophysical observations has been the subject of intense research for quite some time now. There are various models where the LV is incorporated into gravity formulation. The Kalb-Ramond field model and the Bumblebee model are two such models where the Lorentz symmetry is spontaneously broken due to a field's non-zero vacuum expectation value. Our main focus in this endeavor is to conduct a comparative study of the two models concerning the impacts of LV on different observables. We delved into an in-depth study exploring differences in observational signatures of LV arising in KR and BM models. We initiate our inquisition by first introducing the two metrics. The uniqueness of the two solutions is illustrated by their Kretschmann scalars, which are different from the Kretschmann scalar for the \s BH for non-zero KR parameter $\a$ and BM parameter $\b$.\\
The position of the event horizon for a KR BH is $r_h=2M(1-\a)$, indicating a decrease in the horizon radius with $\a$. The horizon radius of a BM BH is the same as that of a \s BH, i.e., $r_h=2M$. We obtain differential equations and the effective potential for null geodesics with the Lagrangian formulation. We get the radius of unstable circular orbits for the two BHs by imposing appropriate conditions on the effective potential and its derivatives. Similar to the horizon radius, increasing $\a$ decreases the photon radius and the associated impact parameter which are given by $r_p=3M(1-\a)$ and $\b_p=3\sqrt{3}(1-\a)^(3/2)$, respectively and their values for a BM BH are same as those for a \s BH i.e., $r_p=3M$ and $b_p=3\sqrt{3}M$. Since the photon radius and the critical impact parameter depend on the nature of the time and azimuthal components of the metric. In contrast, the LV comes into the picture in the radial component of the metric in the BM model, the photon radius and the associated impact parameter do not depend on $\b$. Qualitative and quantitative variations of the photon radius and the critical impact parameter for a KR BH are shown in Fig. [\r{radius}] and Table [\r{shadowradius}]. The importance of the critical impact parameter lies in the fact that a photon with this impact parameter will be trapped in an unstable circular orbit around the BH. The photon on a small perturbation from its circular path may either move towards the BH and never come back or move towards a distant observer, contributing to the optical appearance of the BH. Next, we turn our attention to investigating the shadow of BHs with static and infalling accretions.\\
It is believed that BHs are surrounded by accretion disks. BHs gain mass by accreting ambient matter. Accretion disks we considered are of two types: static and infalling. For both types of accretion, we probed the impact of LV on the observed intensity and the optical appearance of the BH. For static accretion, the following are the conclusions we can make from our study:\\
$(i)$ The observed intensity increases with the LV parameter for both the BHs. However, the intensity peaks at a lower impact parameter for a higher value of KR parameter $\a$, but the position remains the same for BM BH irrespective of the value of $\b$. This can be observed from Figs. [\r{statickr}, \r{staticbm}]. The size of the bright ring decreases with increasing $\a$ but remains the same for BM BH. This is because the intensity peaks at $b_p$ which decreases with $\a$ but remains invariant for BM BH.\\
$(ii)$ The intensity variation with $\a$ is significant for KR BH but is found to be marginal with a variation of $\b$. The brightness of the circular ring in Fig. [\r{statickr}, \r{staticbm}] is indicative of this observation.\\
$(iii)$ The comparison of intensities between KR and BM BHs illustrated in Fig. [\r{staticcom}] reveals that a KR BH will be brighter than a BM BH.\\
The observed intensity in infalling accretion is less, the intensity variation with the impact parameter is smooth, and the central portion of the shadow within the bright ring is darker than the static accretion. The impact of LV parameters $\a$ and $\b$ on the optical appearance with infalling accretion are found to be as follows:
$(i)$ Similar to the static accretion, the intensity for infalling accretion peaks at lower impact parameters for KR BH but remains the same for BM BH due to the reason already stated.\\
$(ii)$ Although the brightness of the photon ring increases with $\a$ similar to the static case, the impact of $\b$ is found to be opposite to that in the static accretion, i.e., the brightness of the photon ring decreases with$\b$. This is because the proper length in BM decreases with $\b$, causing the brightness to fall with increasing $\b$.\\
The above results indicate that the optical appearance with accretion can be used to differentiate between KR and BM BHs based on observables. We then move on to the study of QNMs and ringdown waveforms.\\
When a BH is perturbed due to a field such as scalar or electromagnetic, it returns to equilibrium by emitting GWs. QNMs are these modes that are transient and eventually die down. QNMs are complex-valued functions of model parameters and, as such, provide excellent avenues to probe the impact of additional parameters arising in a model. We use the $6$th order Pad\'{e} averaged WKB method due to its higher accuracy than other WKB methods \c{konoplya2}. We first reduce the Klein-Gordon equation for the massless scalar field and the Maxwell equation for the electromagnetic field to a desired Schr$\ddot{o}$dinger-like equation that yields the potential required to obtain QNMs. We then briefly discuss the $6$th order Pad\'{e} averaged WKB method. We present graphically the variation of frequency of GWs given by the real part of QNMs with LV parameters in Fig. [\r{rekr}, \r{rebm}], and the qualitative variation of the decay rate with LV parameters are shown in Fig. [\r{imkr}, \r{imbm}]. They reveal that the frequency and the decay rate increase with the KR parameter $\a$ for both perturbations. The decay rate decreases with BM parameter $\b$ for scalar and electromagnetic perturbations. However, an increase(decrease) in frequency is observed for an increase in $\b$ for electromagnetic(scalar) perturbation. A comparison of frequency and decay rate between GWs originating from KR and BM BHs is made in Figs. [\r{recom}, \r{imcom}]. It is observed that a KR BH emits GWs of higher frequencies, and they decay faster than those originating from a BM BH. Numerical values given in Tables [\r{QNMS}, \r{QNMEM}] validate these conclusions. We also provide ringdown waveforms for KR and BM BHs. They reveal that a KR BH, once perturbed from its equilibrium, will return to equilibrium quicker than a \s or BM BH by emitting GWs of higher frequency. Comparing a BM BH with a \s BH reveals that a BM BH will take longer to return to equilibrium. \\
A BH emits radiation, and a temperature is associated with the BH horizon. The radiation emitted by a BH at the horizon and the radiation received by a distant observer differ due to the redshift factor. The greybody distribution provides the radiation perceived by an asymptotic observer, and GF gives the probability of Hawking radiation transmission. We intend to observe the impacts of LV on the Hawking temperature, GF, and power emitted. The Hawking temperature is found to be decreasing with $\b$ but increasing with $\a$. As far as GF is concerned, it decreases with $\a$ and $\b$. However, a comparison of GF between two BHs reveals that the probability of the Hawking radiation getting detected by an asymptotic observer is higher for a BM BH. In Figs. [\r{plkr}, \r{plbm}], we illustrate the variation of the power emitted in the form of Hawking radiation with frequency $\omega$ for different values of $\a$ and $\b$. It reveals that the energy emitted by a KR BH is more than that emitted by a BM or \s BH. However, a BM BH emits less energy than a \s BH.\\
Finally, the gravitational lensing in the weak field limit is considered to differentiate between KR and BM BHs based on observational signature. We employed the Gauss-Bonnet theorem to obtain the deflection angle. We obtained higher order correction terms by first obtaining the deflection angle assuming a straight-line trajectory where $r=\frac{b}{sin\phi}$ and then by using that deflection angle and the trajectory given in \cite{CRISNEJO}. Analytical expressions of deflection angles for the KR and BM BHs and graphical representations in Fig. [\r{weak}] reveal that the deflection angle for KR and BM BHs is smaller than that for \s BH. Fig. [\r{defcom}] shows that a light ray deviates more from its straight-line path due to a BM BH than a KR BH. With possible future experimental results on QNMs and deflection angle, we will be able to not only differentiate between KR and BM BHs but also validate or invalidate any of them.


\begin{thebibliography}{the}
\bibitem{Kostelecky1989a}
V.A.~Kostelecky and S.~Samuel, 
\emph{Spontaneous breaking of Lorentz symmetry in string theory}, 
\href{https://doi.org/10.1103/PhysRevD.39.683}
{{Phys. Rev. D} {\bfseries 39} (1989) 683}.

\bibitem{Alfaro2002}
J.~Alfaro, H.A.~{Morales-Tecotl} and L.F.~Urrutia, 
\emph{Loop quantum gravity and light propagation}, 
\href{https://doi.org/10.1103/PhysRevD.65.103509}
{{Phys. Rev. D} {\bfseries 65} (2002) 103509} 
[\href{https://arxiv.org/abs/hep-th/0108061}{{\ttfamily arXiv:hep-th/0108061}}].

\bibitem{Horava2009a}
P.~Horava, 
\emph{Quantum Gravity at a Lifshitz Point}, 
\href{https://doi.org/10.1103/PhysRevD.79.084008}
{{Phys. Rev.} {\bfseries 79} (2009) 084008} 
[\href{https://arxiv.org/abs/0901.3775}{{\ttfamily arXiv:0901.3775}}].

\bibitem{Carroll2001}
S.M.~Carroll, J.A.~Harvey, V.A.~Kostelecky, C.D.~Lane and T.~Okamoto, 
\emph{Noncommutative field theory and Lorentz violation}, 
\href{https://doi.org/10.1103/PhysRevLett.87.141601}
{{Phys. Rev. Lett.} {\bfseries 87} (2001) 141601} 
[\href{https://arxiv.org/abs/hep-th/0105082}{{\ttfamily arXiv:hep-th/0105082}}].

\bibitem{Jacobson2001}
T.~Jacobson and D.~Mattingly, 
\emph{Gravity with a dynamical preferred frame}, 
\href{https://doi.org/10.1103/PhysRevD.64.024028}
{{Phys. Rev. D} {\bfseries 64} (2001) 024028} 
[\href{https://arxiv.org/abs/gr-qc/0007031}{{\ttfamily arXiv:gr-qc/0007031}}].

\bibitem{Dubovsky2005}
S.L.~Dubovsky, P.G.~Tinyakov and I.I.~Tkachev, 
\emph{Massive graviton as a testable cold dark matter candidate}, 
\href{https://doi.org/10.1103/PhysRevLett.94.181102}
{{Phys. Rev. Lett.} {\bfseries 94} (2005) 181102} 
[\href{https://arxiv.org/abs/hep-th/0411158}{{\ttfamily arXiv:hep-th/0411158}}].

\bibitem{Bengochea2009}
G.R.~Bengochea and R.~Ferraro, 
\emph{Dark torsion as the cosmic speed-up}, 
\href{https://doi.org/10.1103/PhysRevD.79.124019}
{{Phys. Rev.} {\bfseries 79} (2009) 124019} 
[\href{https://arxiv.org/abs/0812.1205}{{\ttfamily arXiv:0812.1205}}].

\bibitem{Cohen2006}
A.G.~Cohen and S.L.~Glashow, 
\emph{Very special relativity}, 
\href{https://doi.org/10.1103/PhysRevLett.97.021601}
{{Phys. Rev. Lett.} {\bfseries 97} (2006) 021601} 
[\href{https://arxiv.org/abs/hep-ph/0601236}{{\ttfamily arXiv:hep-ph/0601236}}].

\bibitem{Kostelecky2004a}
V.A.~Kostelecky, 
\emph{Gravity, Lorentz violation, and the standard model}, 
\href{https://doi.org/10.1103/PhysRevD.69.105009}
{{Phys. Rev. D} {\bfseries 69} (2004) 105009} 
[\href{https://arxiv.org/abs/hep-th/0312310}{{\ttfamily arXiv:hep-th/0312310}}].

\bibitem{Kostelecky1989}
V.A.~Kostelecky and S.~Samuel, 
\emph{Gravitational Phenomenology in Higher Dimensional Theories and Strings}, 
\href{https://doi.org/10.1103/PhysRevD.40.1886}
{{Phys. Rev. D} {\bfseries 40} (1989) 1886}.

\bibitem{Kostelecky1989b}
V.A.~Kostelecky and S.~Samuel, 
\emph{Phenomenological gravitational constraints on strings and higher dimensional theories}, 
\href{https://doi.org/10.1103/PhysRevLett.63.224}
{{Phys. Rev. Lett.} {\bfseries 63} (1989) 224}.

\bibitem{Bailey2006}
Q.G.~Bailey and V.A.~Kostelecky, 
\emph{Signals for Lorentz violation in post-Newtonian gravity}, 
\href{https://doi.org/10.1103/PhysRevD.74.045001}
{{Phys. Rev. D} {\bfseries 74} (2006) 045001} 
[\href{https://arxiv.org/abs/gr-qc/0603030}{{\ttfamily arXiv:gr-qc/0603030}}].

\bibitem{Bluhm2008a}
R.~Bluhm, N.L.~Gagne, R.~Potting and A.~Vrublevskis, 
\emph{Constraints and stability in vector theories with spontaneous Lorentz violation}, 
\href{https://doi.org/10.1103/PhysRevD.79.029902}
{{Phys. Rev. D} {\bfseries 77} (2008) 125007} 
[\href{https://arxiv.org/abs/0802.4071}{{\ttfamily arXiv:0802.4071}}].

\bibitem{bm} R. Casana, A. Cavalcante, F. P. Poulis, E. B. Santos, An exact Schwarzschild-like solution in a bumblebee gravity model 	Phys. Rev. D 97, 104001 (2018) [arXiv:1711.02273 [gr-qc]].

\bibitem{Ovgun2018}
A.~Ovg{\"u}n, K.~Jusufi and I.~Sakalli, 
\emph{Gravitational lensing under the effect of Weyl and bumblebee gravities: Applications of Gauss-Bonnet theorem}, 
\href{https://doi.org/10.1016/j.aop.2018.10.012}
{{Annals Phys.} {\bfseries 399} (2018) 193} 
[\href{https://arxiv.org/abs/1805.09431}{{\ttfamily arXiv:1805.09431}}].

\bibitem{Kanzi2019}
S.~Kanzi and {\.I}.~Sakall{\i}, 
\emph{GUP modified hawking radiation in bumblebee gravity}, 
\href{https://doi.org/10.1016/j.nuclphysb.2019.114703}
{{Nucl. Phys. B} {\bfseries 946} (2019) 114703} 
[\href{https://arxiv.org/abs/1905.00477}{{\ttfamily arXiv:1905.00477}}].

\bibitem{Yang2019b}
R.-J.~Yang, H.~Gao, Y.~Zheng and Q.~Wu, 
\emph{Effects of Lorentz breaking on the accretion onto a Schwarzschild-like black hole}, 
\href{https://doi.org/10.1088/0253-6102/71/5/568}
{{Commun. Theor. Phys.} {\bfseries 71} (2019) 568} 
[\href{https://arxiv.org/abs/1809.00605}{{\ttfamily arXiv:1809.00605}}].

\bibitem{Cai2022a}
Z.~Cai and R.-J.~Yang, 
\emph{Accretion of the Vlasov gas onto a Schwarzschild-like black hole},  
\href{https://doi.org/10.1016/j.dark.2023.101292}
{Phys. Dark Univ. {\bfseries 42} (2023) 101292} 
[\href{https://arxiv.org/abs/2205.04826}{{\ttfamily arXiv:2205.04826}}].

\bibitem{Oliveira2021}
R.~Oliveira, D.M.~Dantas and C.A.S.~Almeida, 
\emph{Quasinormal frequencies for a black hole in a bumblebee gravity}, 
\href{https://doi.org/10.1209/0295-5075/ac130c}
{{EPL} {\bfseries 135} (2021) 10003} 
[\href{https://arxiv.org/abs/2105.07956}{{\ttfamily arXiv:2105.07956}}].

\bibitem{Maluf2021}
R.V.~Maluf and J.C.S.~Neves, 
\emph{Black holes with a cosmological constant in bumblebee gravity}, 
\href{https://doi.org/10.1103/PhysRevD.103.044002}
{{Phys. Rev. D} {\bfseries 103} (2021) 044002} 
[\href{https://arxiv.org/abs/2011.12841}{{\ttfamily arXiv:2011.12841}}].

\bibitem{Xu2023}
R.~Xu, D.~Liang and L.~Shao, 
\emph{Static spherical vacuum solutions in the bumblebee gravity model}, 
\href{https://doi.org/10.1103/PhysRevD.107.024011}
{{Phys. Rev. D} {\bfseries 107} (2023) 024011} 
[\href{https://arxiv.org/abs/2209.02209}{{\ttfamily arXiv:2209.02209}}].

\bibitem{Mai2023}
Z.-F.~Mai, R.~Xu, D.~Liang and L.~Shao, 
\emph{Extended thermodynamics of the bumblebee black holes}, 
\href{https://doi.org/10.1103/PhysRevD.108.024004}
{{Phys. Rev. D} {\bfseries 108} (2023) 024004} 
[\href{https://arxiv.org/abs/2304.08030}{{\ttfamily arXiv:2304.08030}}].

\bibitem{Xu2023a}
R.~Xu, D.~Liang and L.~Shao, 
\emph{Bumblebee black holes in light of event horizon telescope observations}, 
\href{https://doi.org/10.3847/1538-4357/acbdfb}
{{Astrophys. J.} {\bfseries 945} (2023) 148} 
[\href{https://arxiv.org/abs/2302.05671}{{\ttfamily arXiv:2302.05671}}].

\bibitem{Liang2023}
D.~Liang, R.~Xu, Z.-F.~Mai and L.~Shao, 
\emph{Probing vector hair of black holes with extreme-mass-ratio inspirals}, 
\href{https://doi.org/10.1103/PhysRevD.107.044053}
{{Phys. Rev. D} {\bfseries 107} (2023) 044053} 
[\href{https://arxiv.org/abs/2212.09346}{{\ttfamily arXiv:2212.09346}}].

\bibitem{Ding2020a}
C.~Ding, C.~Liu, R.~Casana and A.~Cavalcante, 
\emph{Exact Kerr-like solution and its shadow in a gravity model with spontaneous Lorentz symmetry breaking}, 
\href{https://doi.org/10.1140/epjc/s10052-020-7743-y}
{{Eur. Phys. J. C} {\bfseries 80} (2020) 178} 
[\href{https://arxiv.org/abs/1910.02674}{{\ttfamily arXiv:1910.02674}}].

\bibitem{Ding2021a}
C.~Ding and X.~Chen, 
\emph{Slowly rotating Einstein-bumblebee black hole solution and its greybody factor in a Lorentz violation model}, 
\href{https://doi.org/10.1088/1674-1137/abce51}
{{Chin. Phys. C} {\bfseries 45} (2021) 025106} 
[\href{https://arxiv.org/abs/2008.10474}{{\ttfamily arXiv:2008.10474}}].

\bibitem{Wang2022}
H.-M.~Wang and S.-W.~Wei, 
\emph{Shadow cast by Kerr-like black hole in the presence of plasma in Einstein-bumblebee gravity}, 
\href{https://doi.org/10.1140/epjp/s13360-022-02785-6}
{{Eur. Phys. J. Plus} {\bfseries 137} (2022) 571} 
[\href{https://arxiv.org/abs/2106.14602}{{\ttfamily arXiv:2106.14602}}].

\bibitem{Liu2019}
C.~Liu, C.~Ding and J.~Jing, 
\emph{Thin accretion disk around a rotating Kerr-like black hole in Einstein-bumblebee gravity model}, 
\href{https://arxiv.org/abs/1910.13259}{{\ttfamily arXiv:1910.13259}}.

\bibitem{Liu2023}
W.~Liu, X.~Fang, J.~Jing and J.~Wang, 
\emph{QNMs of slowly rotating Einstein-Bumblebee black hole}, 
\href{https://doi.org/10.1140/epjc/s10052-023-11231-5}
{{Eur. Phys. J. C} {\bfseries 83} (2023) 83} 
[\href{https://arxiv.org/abs/2211.03156}{{\ttfamily arXiv:2211.03156}}].

\bibitem{Wang2022a}
Z.~Wang, S.~Chen and J.~Jing, 
\emph{Constraint on parameters of a rotating black hole in Einstein-bumblebee theory by quasi-periodic oscillations}, 
\href{https://doi.org/10.1140/epjc/s10052-022-10475-x}
{{Eur. Phys. J. C} {\bfseries 82} (2022) 528} 
[\href{https://arxiv.org/abs/2112.02895}{{\ttfamily arXiv:2112.02895}}].

\bibitem{Ding2023}
C.~Ding, Y.~Shi, J.~Chen, Y.~Zhou, C.~Liu and Y.~Xiao, 
\emph{Rotating BTZ-like black hole and central charges in Einstein-bumblebee gravity}, 
\href{https://doi.org/10.1140/epjc/s10052-023-11761-y}
{{Eur. Phys. J. C} {\bfseries 83} (2023) 573} 
[\href{https://arxiv.org/abs/2302.01580}{{\ttfamily arXiv:2302.01580}}].

\bibitem{Chen2023}
C.~Chen, Q.~Pan and J.~Jing, 
\emph{Quasinormal modes of a scalar perturbation around a rotating BTZ-like black hole in Einstein-bumblebee gravity}, 
\href{https://doi.org/10.1016/j.physletb.2023.138186}
{{Phys. Lett. B} {\bfseries 846} (2023) 138186} 
[\href{https://arxiv.org/abs/2302.05861}{{\ttfamily arXiv:2302.05861}}].

\bibitem{Gullu2022}
{\.I}.~G{\"u}ll{\"u} and A.~{\"O}vg{\"u}n, 
\emph{Schwarzschild-like black hole with a topological defect in bumblebee gravity}, 
\href{https://doi.org/10.1016/j.aop.2021.168721}
{{Annals Phys.} {\bfseries 436} (2022) 168721} 
[\href{https://arxiv.org/abs/2012.02611}{{\ttfamily arXiv:2012.02611}}].

\bibitem{Zhang2023}
X.~Zhang, M.~Wang and J.~Jing, 
\emph{Quasinormal modes and late time tails of perturbation fields on a Schwarzschild-like black hole with a global monopole in the Einstein-bumblebee theory}, 
\href{https://doi.org/10.1007/s11433-023-2153-6}
{{Sci. China Phys. Mech. Astron.} {\bfseries 66} (2023) 100411} 
[\href{https://arxiv.org/abs/2307.10856}{{\ttfamily arXiv:2307.10856}}].

\bibitem{Lin2023}
R.-H.~Lin, R.~Jiang and X.-H.~Zhai, 
\emph{Quasinormal modes of the spherical bumblebee black holes with a global monopole}, 
\href{https://doi.org/10.1140/epjc/s10052-023-11899-9}
{{Eur. Phys. J. C} {\bfseries 83} (2023) 720} 
[\href{https://arxiv.org/abs/2308.01575}{{\ttfamily arXiv:2308.01575}}].

\bibitem{Ding2021}
C.~Ding, X.~Chen and X.~Fu, 
\emph{Einstein-Gauss-Bonnet gravity coupled to bumblebee field in four dimensional spacetime}, 
\href{https://doi.org/10.1016/j.nuclphysb.2022.115688}
{{Nucl. Phys. B} {\bfseries 975} (2022) 115688} 
[\href{https://arxiv.org/abs/2102.13335}{{\ttfamily arXiv:2102.13335}}].

\bibitem{Jha2021}
S.K.~Jha and A.~Rahaman, 
\emph{Bumblebee gravity with a Kerr-Sen-like solution and its Shadow}, 
\href{https://doi.org/10.1140/epjc/s10052-021-09132-6}
{{Eur. Phys. J. C} {\bfseries 81} (2021) 345} 
[\href{https://arxiv.org/abs/2011.14916}{{\ttfamily arXiv:2011.14916}}].

\bibitem{Ding2023a}
C.~Ding, Y.~Shi, J.~Chen, Y.~Zhou and C.~Liu, 
\emph{High dimensional AdS-like black hole and phase transition in Einstein-bumblebee gravity}, 
\href{https://doi.org/10.1088/1674-1137/aca8f4}
{{Chin. Phys. C} {\bfseries 47} (2023) 045102} 
[\href{https://arxiv.org/abs/2201.06683}{{\ttfamily arXiv:2201.06683}}].

\bibitem{Ovgun2019}
A.~{\"O}vg{\"u}n, K.~Jusufi and {\.I}.~Sakall{\i}, 
\emph{Exact traversable wormhole solution in bumblebee gravity}, 
\href{https://doi.org/10.1103/PhysRevD.99.024042}
{{Phys. Rev. D} {\bfseries 99} (2019) 024042} 
[\href{https://arxiv.org/abs/1804.09911}{{\ttfamily arXiv:1804.09911}}].

\bibitem{Liang2022}
D.~Liang, R.~Xu, X.~Lu and L.~Shao, 
\emph{Polarizations of Gravitational Waves in the Bumblebee Gravity Model},
\href{https://doi.org/10.1103/PhysRevD.106.124019}
{{Phys. Rev. D} {\bfseries 106} (2022) 124019} 
[\href{https://arxiv.org/abs/2207.14423}{{\ttfamily arXiv:2207.14423}}].

\bibitem{Amarilo2023}
K.M.~Amarilo, M.B.F.~Filho, A.A.A.~Filho and J.A.A.S.~Reis, 
\emph{Gravitational waves effects in a Lorentz-violating scenario},
\href{https://arxiv.org/abs/2307.10937}{{\ttfamily arXiv:2307.10937}}.

\bibitem{Altschul2010}
B.~Altschul, Q.G.~Bailey and V.A.~Kostelecky, 
\emph{Lorentz violation with an antisymmetric tensor}, 
\href{https://doi.org/10.1103/PhysRevD.81.065028}
{{Phys. Rev. D} {\bfseries 81} (2010) 065028} 
[\href{https://arxiv.org/abs/0912.4852}{{\ttfamily arXiv:0912.4852}}].

\bibitem{Kalb1974}
M.~Kalb and P.~Ramond, 
\emph{Classical direct interstring action}, 
\href{https://doi.org/10.1103/PhysRevD.9.2273}
{{Phys. Rev. D} {\bfseries 9} (1974) 2273}.

\bibitem{Kao1996}
W.F.~Kao, W.B.~Dai, S.-Y.~Wang, T.-K.~Chyi and S.-Y.~Lin, 
\emph{Induced Einstein-Kalb-Ramond theory and the black hole}, 
\href{https://doi.org/10.1103/PhysRevD.53.2244}
{{Phys. Rev. D} {\bfseries 53} (1996) 2244}.

\bibitem{Kar2003}
S.~Kar, S.~SenGupta and S.~Sur, 
\emph{Static spherisymmetric solutions, gravitational lensing and perihelion precession in Einstein-Kalb-Ramond theory}, 
\href{https://doi.org/10.1103/PhysRevD.67.044005}
{{Phys. Rev. D} {\bfseries 67} (2003) 044005} 
[\href{https://arxiv.org/abs/hep-th/0210176}{{\ttfamily arXiv:hep-th/0210176}}].

\bibitem{Chakraborty2017}
S.~Chakraborty and S.~SenGupta, 
\emph{Strong gravitational lensing \textemdash{} a probe for extra dimensions and Kalb-Ramond field}, 
\href{https://doi.org/10.1088/1475-7516/2017/07/045}
{{JCAP} {\bfseries 07} (2017) 045} 
[\href{https://arxiv.org/abs/1611.06936}{{\ttfamily 1611.06936}}].

\bibitem{Nair2022}
K.K.~Nair and A.M.~Thomas, 
\emph{Kalb-Ramond field-induced cosmological bounce in generalized teleparallel gravity}, 
\href{https://doi.org/10.1103/PhysRevD.105.103505}
{{Phys. Rev. D} {\bfseries 105} (2022) 103505} 
[\href{https://arxiv.org/abs/2112.11945}{{\ttfamily arXiv:2112.11945}}].

\bibitem{Fu2012}
C.-E.~Fu, Y.-X.~Liu, K.~Yang and S.-W.~Wei, 
\emph{Q-form fields on p-branes}, 
\href{https://doi.org/10.1007/JHEP10(2012)060}
{{JHEP} {\bfseries 10} (2012) 060} 
[\href{https://arxiv.org/abs/1207.3152}{{\ttfamily arXiv:1207.3152}}].

\bibitem{Chakraborty2016}
S.~Chakraborty and S.~SenGupta, 
\emph{Solutions on a brane in a bulk spacetime with Kalb-Ramond field}, 
\href{https://doi.org/10.1016/j.aop.2016.01.023}
{{Annals Phys.} {\bfseries 367} (2016) 258} 
[\href{https://arxiv.org/abs/1412.7783}{{\ttfamily 1412.7783}}].

\bibitem{Lessa2020}
L.A.~Lessa, J.E.G.~Silva, R.V.~Maluf and C.A.S.~Almeida, 
\emph{Modified black hole solution with a background Kalb-Ramond field}, 
\href{https://doi.org/10.1140/epjc/s10052-020-7902-1}
{{Eur. Phys. J. C} {\bfseries 80} (2020) 335} 
[\href{https://arxiv.org/abs/1911.10296}{{\ttfamily arXiv:1911.10296}}].

\bibitem{Atamurotov2022}
F.~Atamurotov, D.~Ortiqboev, A.~Abdujabbarov and G.~Mustafa, 
\emph{Particle dynamics and gravitational weak lensing around black hole in the Kalb-Ramond gravity}, 
\href{https://doi.org/10.1140/epjc/s10052-022-10619-z}
{{Eur. Phys. J. C} {\bfseries 82} (2022) 659}.

\bibitem{Kumar2020c}
R.~Kumar, S.G.~Ghosh and A.~Wang, 
\emph{Gravitational deflection of light and shadow cast by rotating Kalb-Ramond black holes}, 
\href{https://doi.org/10.1103/PhysRevD.101.104001}
{{Phys. Rev. D} {\bfseries 101} (2020) 104001} 
[\href{https://arxiv.org/abs/2001.00460}{{\ttfamily arXiv:2001.00460}}].

\bibitem{Lessa2021}
L.A.~Lessa, R.~Oliveira, J.E.G.~Silva and C.A.S.~Almeida, 
\emph{Traversable wormhole solution with a background Kalb-Ramond field}, 
\href{https://doi.org/10.1016/j.aop.2021.168604}
{{Annals Phys.} {\bfseries 433} (2021) 168604} 
[\href{https://arxiv.org/abs/2010.05298}{{\ttfamily arXiv:2010.05298}}].

\bibitem{Maluf2022}
R.V.~Maluf and C.R.~Muniz, 
\emph{Exact solution for a traversable wormhole in a curvature-coupled antisymmetric background field}, 
\href{https://doi.org/10.1140/epjc/s10052-022-10409-7}
{{Eur. Phys. J. C} {\bfseries 82} (2022) 445} 
[\href{https://arxiv.org/abs/2110.12202}{{\ttfamily 2110.12202}}].

\bibitem{Maluf2022a}
R.V.~Maluf and J.C.S.~Neves, 
\emph{Bianchi type I cosmology with a Kalb-Ramond background field}, 
\href{https://doi.org/10.1140/epjc/s10052-022-10109-2}
{{Eur. Phys. J. C} {\bfseries 82} (2022) 135} 
[\href{https://arxiv.org/abs/2111.13165}{{\ttfamily arXiv:2111.13165}}].
\bibitem{kr} Ke Yang, Yue-Zhe Chen, Zheng-Qiao Duan, Ju-Ying Zhao, Static and spherically symmetric black holes in gravity with a background Kalb-Ramond field: Phys. Rev. D 108, 124004 (2023) [arXiv:2308.06613 [gr-qc]].
\bibitem{IG} Ignacio Gonzlez Martnez-Pas, Tariq Shahbaz, Jorge Casares Velzquez, Accretion Processes in Astrophysics, Cambridge University Press, 2014.
\bibitem{RN} R. Narayan, M. D. Johnson and C. F. Gammie, Astrophys. J. 885, no. 2, L33 (2019)
\bibitem{HF} H. Falcke, F. Melia and E. Agol, Astrophys. J. Lett. 528 (2000), L13
\bibitem{CB} C. Bambi, Phys. Rev. D 87 (2013), 107501
\bibitem{RS} R. Shaikh and P. S. Joshi, JCAP 10 (2019), 064
\bibitem{KJ} K. Jusufi and Saurabh, Mon. Not. Roy. Astron. Soc. 503 (2021), 1310
\bibitem{KS} K. Saurabh and K. Jusufi, Eur. Phys. J. C 81 (2021) no.6, 490
\bibitem{SN} S. Nampalliwar, S. Kumar, K. Jusufi, Q.Wu, M. Jamil and P. Salucci, Astrophys. J. 916 (2021) no.2, 116
\bibitem{XX} X. X. Zeng, H. Q. Zhang and H. Zhang, [arXiv:2004.12074].
\bibitem{KONOR} R. A. Konoplya and A. Zhidenko, Rev. Mod. Phys. 83, 793 (2011)
\bibitem{PARTERBATION1} T. Regge, J. A. Wheeler:  Phys. Rev., 108:1063-1069, (1957)
\bibitem{PARTERBATION2} F. J. Zerilli:
Phys. Rev. Lett., 24:737-738, (1970)
\bibitem{PARTERBATION3} F. J. Zerilli.  Phys. Rev., D2:2141-2160, (1970)
\bibitem{PARTERBATION4} F. J.
Zerilli:  Phys. Rev., D9:860-868, (1974)
\bibitem{PARTERBATION5} V. Moncrief: Phys. Rev., D12:1526-1537,
(1975)
\bibitem{PARTERBATION6} S. A. Teukolsky: Phys. Rev. Lett., 29:1114-1118,
(1972)
\bibitem{LIGO1}
P.B. Abbott \textit{et al.} [LIGO1 Scientific and Virgo]
Astrophysical Implications of the Binary Black-Hole Merger
GW150914
 \emph{Astrophys. J. Lett.} \textbf{818} 2 L22 (2016)

\bibitem{LIGO2}
P.B. Abbott \textit{et al.} [LIGO Scientific and Virgo]
Observation of Gravitational Waves from a Binary Black Hole Merger
 \emph{Phys. Rev. Lett.} \textbf{116} 6 061102 (2016)

\bibitem{LIGO3}
P.B. Abbott \textit{et al.} [LIGO Scientific and Virgo] 2016
Binary Black Hole Mergers in the first Advanced LIGO Observing Run
 \emph{Phys. Rev. X} \textbf{6} 4 041015 (2016)

\bibitem{LIGO4}
P.B. Abbott \textit{et al.} [LIGO Scientific and VIRGO] GW170104:
Observation of a 50-Solar-Mass Binary Black Hole Coalescence at
Redshift 0.2
 \emph{Phys. Rev. Lett.} \textbf{118} 22 221101 (2017)
\bibitem{VIRGO} F. Acernese \textit{et al.}
(Virgo Collaboration),  Physical Review Letters. 123 (23): 231108
(2019)
\bibitem{13}J. S. F. Chan and R. B. Mann, Phys. Rev. D 55  7546
(1997)
\bibitem{14} G.T. Horowitz , V. E. Hubeny, Phys. Rev. D62
024027  (2000
\bibitem{15} S. Hod, Phys. Rev. Lett. 81  4293 (1998)
\bibitem{16} R. A. Konoplya and A. Zhidenko, Rev. Mod. Phys. 83
793  (2011)
\bibitem{17} B. Chen and J. Zhang, Phys. Rev. 84  124039 (2011)
arXiv:1110.3991 [hep-th]
\bibitem{18} Y. Kim, Y. S. Myung and Y. Park, Eur. Phys. J. C 73 138 (2013)
\bibitem{19} I. Z. Stefanov, S. S. Yazadjiev and G. G. Gyulchev,  Phys. Rev. Lett. 104
251103 (2010)
\bibitem{20} R. A. Konoplya, A. F. Zinhailo,  Z. StuchlikPhys. Rev. D 102, 044023 (2020)
\bibitem{KONO} R. A. Konoplya and A. Zhidenko, Phys. Rev. D 105, 104032 (2022)
\bibitem{KONO1} K. A. Bronnikov, R. A. Konoplya, and T. D. Pappas, Phys. Rev. D 103, 124062 (2021)
\bibitem{KONO2} R. A. Konoplya, Phys. Rev. D 103, 044033 (2021).
\bibitem{21} T. V. Fernandes, D. Hilditch, J. P. S. Lemos,  V. Cardoso, Phys. Rev. D 105 044017 (2022)
\bibitem{22} K. Jusu, M. Azreg-Anou, M. Jamil, Shao-Wen Wei, Q. Wu,  A. Wang, Phys. Rev. D 103, 024013 (2021)
 \bibitem{23} E. Franzin, S. Liberati, J. Mazza, R. Dey, S. Chakraborty, Phys. Rev. D 105 124051 (2022)
\bibitem{24} S. Chakraborty, K. Chakravarti, S. Bose, S. SenGupta, Phys. Rev. D 97, 104053 (2018)
\bibitem{25} Q. Tan, Wen-Di Guoab, Yu-Xiao Liu, Phys. Rev. D 106, 044038 (2022)
\bibitem{27} P. H. C. Siqueira, M. Richartz, Phys. Rev. D 106, 024046 (2022)
\bibitem{28} T. Torres, S. Patrick, M. Richartz,  S. Weinfurtner, Phys. Rev. Lett. 125, 011301 (2020),
\bibitem{29} T. Assumpcao, V. Cardoso, A. Ishibashi, M. Richartz,  M. Zilhao, Phys. Rev. D 98, 064036 (2018)
\bibitem{30} M. Richartz, Phys. Rev. D 93, 064062 (2016)
\bibitem{31} Wei-Liang Qian , K. Lin, Xiao-Mei Kuang, B. Wang Rui-Hong Yue, Eur. Phys. J. C. 82, 188 (2022),
\bibitem{32} W. Yao, S. Chen, and J. Jing, Phys. Rev. D 83, 124018 (2011)
\bibitem{33} M. Okyay, A. $\ddot{O}$vg $\ddot{u}$n, JCAP 01, 009 (2022)
\bibitem{34} D. Liu, Y. Yang, S. Wu, Y. Xing, Z. Xu,  Z.-W. Long, Phys. Rev. D 104, 104042 (2021)
\bibitem{35} G. Guo, P. Wang, H. Wu, H. Yang, JHEP 06, 060 (2022), arXiv:2112.14133 [gr-qc]
\bibitem{36} M. S. Churilova, R. A. Konoplya,  A. Zhidenko, Phys. Lett. B 802, 135207 (2020)
\bibitem{37} V. Cardoso, S. Hopper, C. F. B. Macedo, C. Palenzuela, P. Pani, Phys. Rev. D 94, 084031 (2016)
\bibitem{38} V. Cardoso, V. F. Foit,  M. Kleban, JCAP 08, 006 (2019)
\bibitem{39} M. R. Correia, V. Cardoso, Phys. Rev. D 97, 084030 (2018)
\bibitem{40} R. A. Konoplya, A. Zhidenko, EPL 138, 49001 (2022)
\bibitem{41} M. S. Churilova, R. A. Konoplya, Z. Stuchlik, A. Zhidenko, JCAP 10, 010 (2021)
\bibitem{42} K. A. Bronnikov, R. A. Konoplya, Phys. Rev. D 101, 064004 (2020)
\bibitem{43} R. A. Konoplya, Z. Stuchlk,  A. Zhidenko, Phys. Rev. D 99, 024007 (2019)
\bibitem{44} V. F. Foit, M. Kleban, Class. Quant. Grav. 36, 035006 (2019)
\bibitem{45} Yu-Tong Wang, Zhi-Peng, J. Zhang, Shuang-Yong Zhou, Yun-Song Piao, Eur. Phys. J. C 78, 482
 (2018)
\bibitem{47} P. Pani,  V. Ferrari, Class. Quant. Grav. 35, 15LT01 (2018)
\bibitem{48} A. Testa, P. Pani, Phys. Rev. D 98, 044018 (2018)
\bibitem{49} E. Maggio, A. Testa, S. Bhagwat,  P. Pani, Phys. Rev. D 100, 064056 (2019)
\bibitem{50} N. Oshita and N. Afshordi, Phys. Rev. D 99, 044002 (2019)
\bibitem{mj97}
M. Jaroszynski and A. Kurpiewski, Optics near kerr black holes: spectra of advection dominated accretion flows, Astron. Astrophys. \textbf{326}: 419 (1997).
\bibitem{bambi13}
C. Bambi, Can the supermassive objects at the centers of galaxies be traversable wormholes? The first test of strong gravity for mm/sub-mm very long baseline interferometry facilities. Phys. Rev. D. \textbf{87}: 107501 (2013).
\bibitem{schutz} F. B. Schutz  M. C.  Will  Black hole normal modeds:
A schematic approach Astrophys. J. Lett. 291 L33-L36 (1985)
\bibitem{iyer} S. Iyer M. C. Will Black Hole Normal Modes: A {WKB} Approach.
1. Foundations and Application of a Higher Order {WKB} Analysis of Potential
 Barrier Scattering Phys. Rev. D 35 3621 (1987)
\bibitem{iyer1} S. Iyer  Black hole normal modeds:
A WKB aproach 2. Schwarzschild black holes Phys. Rev. D 35  3632
(1987)
\bibitem{konoplya1} R. Konoplya  Quasinormal behavior of the d-dimensional
Schwarzschild black hole and higher order WKB approach Phys. Rev.
D68  024018 (2003)
\bibitem{jerzy}Jerzy Matyjasek and Micha l Opala, “Quasinormal modes of black holes: The improved semianalytic approach,”
Phys. Rev. D 96, 024011 (2017).
\bibitem{konoplya2} R. A. Konoplya, A. Zhidenko, A. F. Zinhailo, Class. Quantum Grav. 36: 155002 (2019) [arXiv:1904.10333].
\bibitem{gundlach1} C. Gundlach , H. R. Price, J. Pullin: \emph{Phys. Rev. D} \textbf{49} 890-899 (1994).
\bibitem{HAWKING}S. W. Hawking, Commun. Math. Phys. 43, 199 (1975b), [Erratum: Commun.Math.Phys. 46, 206 (1976)]
\bibitem{HH} H. Hassanabadi et al., “Effects of a new extended uncertainty principle on Schwarzschild and Reissner–Nordstr¨om black holes thermodynamics,” Int. J. Mod. Phys. A 36, 2150036 (2021).
\bibitem{SH} S. Hassanabadi et  al., “Thermodynamics of the Schwarzschild and Reissner–Nordstr¨om black holes under higher-order generalized uncertainty principle,” Eur. Phys. J. Plus 136, 918 (2021),
arXiv:2110.01363 [gr-qc].
\bibitem{HC} Hao Chen et  al., “Thermodynamics of the Reissner-Nordstr¨om black hole with quintessence matter on the EGUP framework,” Phys. Lett. B 827, 136994 (2022).
\bibitem{BEK} Black hole thermodynamics, Physics Today 24, Bekenstein 1980.
\bibitem{KEIF} Classical and Quantum black holes by Keifer 1999.
\bibitem{SW} Shao-Wen Wei Zhang, Yu-Peng and Yu-Xiao Liu, “Topological approach to derive the global Hawking temperature of (massive) BTZ black hole.” Physics Lett. B 810 (2020).
\bibitem{ALI} Ali $\ddot{O}$vg$\ddot{u}$n and Izzat Sakalli, “Hawking radiation via gaussbonnet theorem,” Ann. of Phys. 413, 168071 (2020).
\bibitem{SI} S. I Kruglov, “Magnetically charged black hole in framework of nonlinear electrodynamics model.” Int. J. of Mod. Phys. A 33 (2018).
\bibitem{qn32} Creek, S., Efthimiou, O.; Kanti, P. and Tamvakis, K. Phys. Rev. D, 76, 104013 (2007).

\bibitem{qn36} Shankaranarayanan, S. Phys. Rev. D, 67, 084026 (2003).
\bibitem{qn37} Boonserm, P. and Visser, M. . Ann. Phys., 325, 1328--1339 (2010).

\bibitem{qn38} Kanzi, S., Sakall\i , I. Nucl. Phys. B, 946, 114703 (2019).

\bibitem{qn39} Al-Badawi, A., Sakall\i , I. and Kanzi, S. Ann. Phys. 2020, 412, 168026.

\bibitem{qn40} Al-Badawi, A., Kanzi, S. and Sakall\i , I. Eur. Phys. J. Plus 2020, 135,
219.

\bibitem{qn41} P. Boonserm, T. Ngampitipan and P. Wongjun, Eur. Phys. J. C 79, 330 (2019), arXiv:1902.05215 [gr-qc].

\bibitem{qn42} M. Visser, Phys. Rev. A 59, 427 (1999), arXiv:quant-ph/9901030.

\bibitem{qn43} C. V. Vishveshwara, Nature 227, 936 (1970).

\bibitem{qn44} K. D. Kokkotas and B. G. Schmidt, Living Rev. Rel. 2, 2 (1999),
arXiv:gr-qc/9909058.

\bibitem{qn45} H.-P. Nollert, Class. Quant. Grav. 16, R159 (1999).

\bibitem{qn46} V. Cardoso and P. Pani, Living Rev. Rel. 22, 4 (2019), arXiv:1904.05363
[gr-qc].

\bibitem{GB}Matt Visser, “Some general bounds for one-dimensional scattering,” Phys. Rev. A 59, 427–438 (1999).
\bibitem{GB1}Petarpa Boonserm and Matt Visser, “Bounding the bogoliubov coefficients,” Annals of Physics 323, 2779–2798 (2008).
\bibitem{GB2}Boonserm. P, “Rigorous bounds on transmission, reflection and bogoliubov coefficients,” Ph.D. thesis, Victoria Uni. Wellington (2009).
\bibitem{WJ} W. Javed, I. Hussain, and A. O¨ vgu¨n, “”Weak deflection angle of KazakovSolodukhin black hole in plasma medium using GaussBonnet theorem and its greybody bonding.” Eur. Phys. J. Plus 137 (2022).
\bibitem{yg2017}Y.-G. Miao and Z.-M. Xu, Hawking Radiation of Five-Dimensional Charged Black Holes with Scalar
Fields, Phys. Lett. B 772, 542 (2017).
\bibitem{fg2016} F. Gray, S. Schuster, A. Van–Brunt, and M. Visser, The Hawking Cascade from a Black Hole Is
Extremely Sparse, Class. Quantum Grav. 33, 115003 (2016).
\bibitem{251}Hoekstra, Henk and others, “Masses of galaxy clusters from gravitational lensing.” Space Sci. Rev. 177, 75–118 (2013).
\bibitem{261} Brouwer, M. M. and others, “Studying galaxy troughs and ridges using weak gravitational lensing with the Kilo-Degree Survey.” , Mon.Not. Roy. Astron. Soc 481, 5189 (2018).
\bibitem{271} R. Ali Vanderveld, Michael J. Mortonson, Wayne Hu, and Tim Eifler, “Testing dark energy paradigms with weak gravitational lensing,”Phys. Rev. D 85, 103518 (2012).
\bibitem{GW} G. W. Gibbons and M. C. Werner, Class. Quant. Grav. 25, 235009 (2008).
\bibitem{CARMO}M. P. Do Carmo, Differential Geometry of Curves and Surfaces, (Prentice-Hall, New Jersey, 1976).
\bibitem{WERNER} M. C. Werner, Gen. Rel. Grav. 44, 3047 (2012).
\bibitem{ISHIHARA1} A. Ishihara, Y. Suzuki, T. Ono, T. Kitamura, H. Asada. Phys. Rev., D94(8):084015, (2016).
\bibitem{ISHIHARA2} A. Ishihara, Y. Suzuki, T. Ono and H. Asada, Phys. Rev. D 95, 044017 (2017).
\bibitem{ONO1}T. Ono, A. Ishihara and H. Asada, Phys. Rev. D 96, 104037 (2017).
\bibitem{ONO2} T. Ono, A. Ishihara, H. Asada: Phys. Rev., D96(10):104037, (2017).
\bibitem{ONO3} T. Ono, A. Ishihara, H. Asada. Phys. Rev. D98(4):044047, (2018).
\bibitem{CRISNEJO} G. Crisnejo, E. Gallo and K. Jusufi, Phys. Rev. D100, 104045 (2019).
\bibitem{411}Zonghai Li and Ali A.~Ovg{\"u}n, “Finite-distance gravitational deflection of massive particles by a kerr-like black hole in the bumblebee gravity model,” Phys. Rev. D 101, 024040 (2020)
\bibitem{421} Zonghai Li, Guodong Zhang, and Ali A.~Ovg{\"u}n, “Circular orbit of a particle and weak gravitational lensing,” Phys. Rev. D 101, 124058 (2020).
\bibitem{431} J. H. Oort, “The force exerted by the stellar system in the direction perpendicular to the galactic plane and some related problems,” Astron. Inst. Netherlands 6, 249 (1932).
\bibitem{441} Fritz. Zwicky, “On the masses of nebulae and of clusters of nebulae.” The Astrophys. J. 86, 217 (1937).
\bibitem{451} J. L. Feng, “Dark Matter Candidates from Particle Physics and Methods of Detection.” The Astrophys. J. Supple. Ser. 48, 495–545 (2010).
\bibitem{461} N. Jarosik et al., Astrophys. J., Suppl. Ser. 192, 14 (2011).
\bibitem{471} A. A.~Ovg{\"u}n, “Deflection angle of photons through dark matter by black holes and wormholes using gaussbonnet theorem,” Universe 5, 115 (2019).
\bibitem{481} Reggie C. Pantig and A.~Ovg{\"u}n, “Dark matter effect on the weak deflection angle by black holes at the center of Milky Way and M87 galaxies,” Eur. Phys. J. C 82, 391 (2022), arXiv:2201.03365 [gr-qc].
\bibitem{491} Reggie C. Pantig and Emmanuel T. Rodulfo, “Weak deflection angle of a dirty black hole,” Chin. J. Phys. 66, 691–702 (2020).
\bibitem{501} Reggie C. Pantig and A.~Ovg{\"u}n, “Black hole in quantum wave dark matter,” Fortsch. Phys. 2022, 2200164 (2022), arXiv:2210.00523 [gr-qc].
\bibitem{511}Reggie C. Pantig and A.~Ovg{\"u}n, “Dehnen halo effect on a black hole in an ultra-faint dwarf galaxy,” JCAP 08, 056 (2022), arXiv:2202.07404 [astro-ph.GA].


\end{thebibliography}
\end{document}